\documentclass{aa}
\usepackage{epsfig}
\usepackage{amsmath}
\usepackage{times}                       % closer to printed appearance
\usepackage{graphicx}                    % Figures easier to handle
  \newcommand{\Msolar}{\mbox{\,$\rm M_{\odot}$}}        % solar mass
  \newcommand{\Lsolar}{\mbox{\,$\rm L_{\odot}$}}        % solar luminosity
  \newcommand{\Teff}{\mbox{\,\em T$_{\rm eff}$}}         % effective temperature
  \newcommand{\te}{\mbox{\,\em T$_{\rm eff}$}}           % effective temperature
  \newcommand{\sg}{\mbox{\,log $g$}}                     % surface gravity
  \newcommand{\iso}[2]{\mbox{$^{#1}{\rm #2}$}}           % isotope designation
  \newcommand{\Wl}{\mbox{\,$W_{\lambda}$}}               % equivalent width
  \newcommand{\vt}{\mbox{\,$v_{\rm t}$}}                 % microturbulence
  \newcommand{\Ebv}{\mbox{\,$E_{\rm B-V}$}}              % extinction
  \newcommand{\Eub}{\mbox{\,$E_{\rm U-B}$}}              % extinction
  \newcommand{\nH}{\mbox{\,$n_{\rm H}$}}                 % hydrogen abundance
  \newcommand{\nHe}{\mbox{\,$n_{\rm He}$}}               % helium abundance
  \newcommand{\kmsec}{\,\mbox{$\mbox{km}\,\mbox{s}^{-1}$}}    % kilometres/second
  \newcommand{\kelvin}{\,\mbox{K}}                       % K Kelvin
  \def\simge{\mathrel{\raise1.16pt\hbox{$>$}\kern-7.0pt
    \lower3.06pt\hbox{{$\scriptstyle \sim$}}}}           % approx ge
  \def\simle{\mathrel{\raise1.16pt\hbox{$<$}\kern-7.0pt
    \lower3.06pt\hbox{{$\scriptstyle \sim$}}}}           % approx le
\begin{document}

\title{Stellar archaeology: the evolving spectrum of FG\,Sge} 

\titlerunning{Spectral evolution of FG\,Sge}

   \author{C.S. Jeffery \inst{1}
          \and  D. Sch\"onberner\inst{2} 
          }
         
   \authorrunning{ C.S. Jeffery \& D. Sch\"onberner }
   
   \offprints{C.S. Jeffery, \email{csj@arm.ac.uk}}
   \institute{Armagh Observatory, College Hill, 
            Armagh BT61 9DG, Northern Ireland. 
   \and Astrophysikalisches Institut, Potsdam, 
        An der Sternwarte 16, D-14482 Potsdam, Germany.}
   \date{Received; accepted \ldots }

%
% _____________________________________________________________________ abstract
%
\abstract{Over an interval of 120 years, the extraordinary 
object \object{FG\,Sge} has been transformed from a hot post-AGB star to
a very luminous cool supergiant. Theoretically, this evolution has been 
associated with the reignition of a helium-shell during the post-AGB stage. 
A series of studies of the chemical composition of the photosphere have
suggested that this evolution has been associated with a dramatic 
increase of approximately 3 dex in the abundances of s-process
elements between about 1960 and 1995. The problem with this apparent 
change is that it occurred at a time when the surface convection
zone, which is governed by the star's effective temperature, 
could not have developed sufficiently to dredge processed material from
the stellar interior to the surface. We have reviewed
the chemical evolution of \object{FG\,Sge} by means of modelling the
time-varying spectrum under a range of assumptions. By comparing
these models with published observational data, a self-consistent
picture has emerged. In particular we find that surface 
hydrogen has been depleted during the interval in question. 
In contrast, the s-process abundances have generally
maintained a steady enhancement of around $1 - 2$ dex, although 
some modest changes may have occurred since 1950.
This implies that \object{FG\,Sge} has not just
completed dredging up freshly-produced s-process
isotopes. {However, there remains a contradiction between the 
observed hydrogen-depletion, the age of the associated planetary
nebula, and 
current evolutionary models for a pre-white dwarf suffering a late thermal
pulse.}
}

\maketitle

% ________________________________________________________________ introduction
%
\section{Introduction}

During the last century, \object{FG\,Sagittae} evolved from a hot central
star of a planetary nebulae to become an F-type supergiant within 
approximately 90 years. Only two comparable objects are known;
\object{V4334\,Sgr} (Sakurai's object) 
covered a similar evolution in approximately five years
to become, in 1999,  a C-type giant smothered by its own dust 
shell (Duerbeck et al.\ \cite{Due02}),
and \object{V605\,Aql} which, after a brief foray into the 
giant domain during the
1920's, returned to the high-temperature r\'egime
as a typical [WC]-type planetary nebula central star (Clayton \& de
Marco \cite{Cla97}). The rarity
and brevity of such events arising from a 
cataclysm in the final evolution of low-mass stars is such that
they are of profound interest.

For these reasons, every observation of these stars is 
important, not so much for their contemporary significance, but for
their contribution to retrospective studies of evolutionary 
episodes which, even for such astronomically rapid events, may
last longer than most scientists' careers. In particular, the
development of increasingly sophisticated analytical methods
render the original data of primary importance in any historical
interpretation. 

The generally accepted interpretation for the redward evolution of
 \object{FG Sge} is that of a late thermal pulse
in a young white dwarf or planetary nebula central star 
(Paczynski \cite{Pac71}; Langer, Kraft \& Anderson \cite{Lan74}; 
Sackmann \& Despain \cite{Sac74}; 
Sch\"onberner \cite{Sch79};
Iben  \cite{Ibe84}; 
Bl\"ocker \& Sch\"onberner \cite{Blo97}; 
Gonzalez et al. \cite{Gon98}; 
Lawlor \& MacDonald \cite{Law03}) 
wherein nuclear helium-burning
at the surface of the electron-degenerate carbon-oxygen core
is reignited. Detailed models for these late
thermal pulses have been generated by several investigators
(cf.\ Sch\"onberner \cite{Sch79}; Iben \cite{Ibe84}; 
Bl\"ocker \cite{Blo95}), but the main 
features are principally the same.
The energy released by the helium-shell flash forces the outer layers
to expand in about 50 to a few 100 years, depending on their mass. 
Hydrogen-rich, helium-rich and carbon-rich material is mixed
by convective overshoot from the helium-shell flash, but the 
surface of the star retains
its pre-pulse composition. A major component of this
mixed region will be a supply of newly-produced  
s-process elements generated by the mixing of \iso{13}{C}
and protons.  As the star expands, its surface cools
allowing surface convection to occur. As the temperature 
bottoms-out, the convection zone digs deeper, dredging processed
material to the stellar surface.

The evolution of \object{FG Sge} in terms of effective
temperature (\Teff) and luminosity ($L$) has been established
from long-term photometric studies (van Genderen \& Gautschy \cite{vGG95}).
These data appear to support the late thermal pulse hypothesis 
(Bl\"ocker \& Sch\"onberner \cite{Blo97}; Herwig \cite{Her01a}; 
Lawlor \& MacDonald \cite{Law03}).
Further evidence is the apparent enhancement
of s-process elements in the photosphere of \object{FG\,Sge} 
(Langer et al.\ \cite{Lan74}) as
\Teff\ dropped below 8\,500\kelvin, almost
exactly the point where surface convection is expected to 
develop (cf.\ van Genderen \& Gautschy \cite{vGG95}). The dilemma, in simple terms, 
is that the theoretical models for late thermal pulse evolution 
do not predict the dredge-up of s-processed material to the 
surface until some considerable time after the star has reached
its minimum effective temperature. Moreover, the dredge-up
should also be associated with a significant change in the
abundances of hydrogen, helium, carbon and oxygen. 
Iben (\cite{Ibe84}) remarks that {\it ``The heretical conclusion 
to which one appears to be forced is that FG\,Sge has {\it not} just
completed dredging up freshly produced s-process isotopes. If this is
so, then the fault must lie in the estimates of surface abundances.''}

Consequently, either the theoretical predictions or the
observations concerning the surface abundance of late thermal
pulse products is wrong, or \object{FG\,Sge} is not a late thermal
pulse product.  

More recently, a variant of the late 
thermal pulse phenomenon has been identified (Herwig et
al. \cite{Her99})\footnote{In fact, both classes of thermal pulse were
 previously encountered, but not explicitly labelled, by Sch\"onberner
 (\cite{Sch79}). }. 
The ``late thermal pulse'' (LTP) model introduced above does 
qualitatively well in reproducing
the hydrogen-deficiency and high carbon abundances seen in 
hydrogen-deficient planetary nebulae central stars, but fails to
reproduce the large oxygen abundances found therein. 
Herwig et al. (\cite{Her99}) found, however, that a ``very late thermal
pulse'' (VLTP), occurring after the H-burning shell has become inactive, 
was able to produce a  surface
abudance pattern in good general agreement with that observed in the 
hydrogen-deficient post-AGB stars, primarily because he considered
convective overshoot.
The two models are different in one major respect. In the first,
the convection zone produced in the He-rich shell at He ignition
cannot extend into the hydrogen-rich envelope (Iben \cite{Ibe76}). In
the second, this convection zone is able to penetrate into the H-rich
surface layers and transport protons downwards (Herwig et al. \cite{Her99}).
What is not immediately apparent from the published figures, although it is
stated in the text, is that the nuclear-driven convection zones are able to
propagate through to the surface before the star cools sufficiently to
develop a deep opacity-driven convection layer at its surface.  
This means that in the VLTP model, the disappearance of surface
hydrogen and the appearance of highly-processed material can occur
quickly and while the effective temperature is still significantly
high. 

Another major difference between the LTP and VLTP models is that the latter
occurs on a much shorter timescale than the former. A VLTP occurs
when the envelope mass is close to $\sim 10^{-4}$\Msolar. Since 
the nuclear energy produced by protons
ingested into the intershell region is enormous ($\sim10^{39}$ erg in
less than a week), expansion following a VLTP occurs on the very small 
thermal timescale ($\sim1$ y) of the tiny hydrogen-rich envelope 
(Herwig et al. \cite{Her97}). 
In contrast, a LTP occurs when the envelope mass is also
$\sim 10^{-4}$\Msolar, 
but $\sim 10^{-2}$\Msolar of hydrogen-free intershell must 
also be lifted, there is no proton-ingestion, and
the expansion energy comes from the He-shell flash alone.
Consequently the expansion occurs
on a much longer timescale ($\sim100$ y). 

The VLTP model is therefore important for understanding the
observed abundance evolution of FG\,Sge and V4334\,Sgr 
(Gonzalez et al. \cite{Gon98}, Asplund et al. \cite{Asp97}). 
For example, it has been stated by several authors that, 
because of their respective evolution timescales, 
FG\,Sge and V4334\,Sgr are examples of LTP and 
VLTP evolution respectively ({\it cf.} Herwig \cite{Her01a}).

It is the intent of this investigation to
determine whether observation can indeed be reconciled with theory, 
and whether FG\,Sge has suffered either a ``late'' or 
a ``very late'' thermal pulse.  
The next sections give a more detailed discussion of 
specific observations, followed by a presentation of the
model atmospheres and spectra used and our subsequent analysis. 
It considers in more detail ideas introduced by Sch\"onberner \& Jeffery (2003). 

The literature on \object{FG\,Sge} is extensive, there
are currently over 394 citations in the {\sc simbad} database
between 1960 and 
the present (2006 July)\footnote{There is only one citation in
  SIMBAD from
  before 1960: Hoffmeister's (\cite{Hof44}) list of 171 new
  variables.}. 
It would be impossible and inappropriate to survey critically
every one of these articles, but inadvertent omissions of 
material that others might consider seminal will no doubt have been
made. However we recognise the contribution of all of these 
publications, and hence our prefatory
remarks on the importance of primary data. These derive from
the fact that much early spectroscopic material is no longer
accessible to us. Consequently, 
any conclusions which run contrary
to the original authors' are legitimately disputable. 
On the other hand, if it can be shown that a self-consistent picture
of \object{FG\,Sge} could be realised by a reassessment of these
data, this paper will have achieved its goal.

%______________________________________________________________________________
% figure showing g-Teff evolution from different sources
\begin{figure}
\epsfig{file=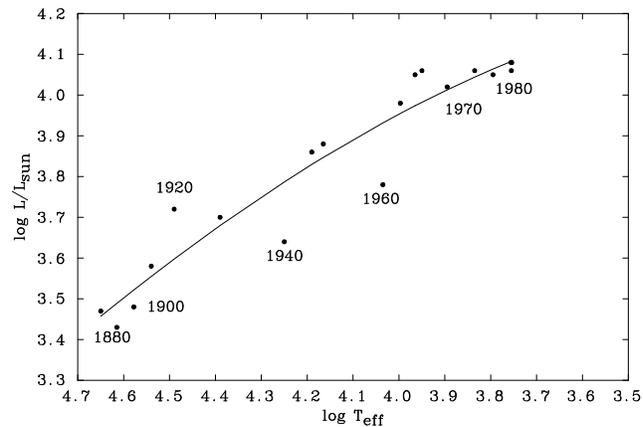,width=85mm} % ,bbllx=160,bblly=15,bburx=350,bbury=350}
\caption[\object{FG\,Sge} evolution]{The evolution of \object{FG\,Sge} 
in the $L$--\Teff\ diagram (van Genderen \& Gautschy \cite{vGG95}, Table 3: high \Teff ). A smooth curve has
been drawn through the data.}
\label{f:evol}
\end{figure}

%______________________________________________________________________________
% Figure "teff"   Observed Teff measurements
%
\begin{figure}
\epsfig{file=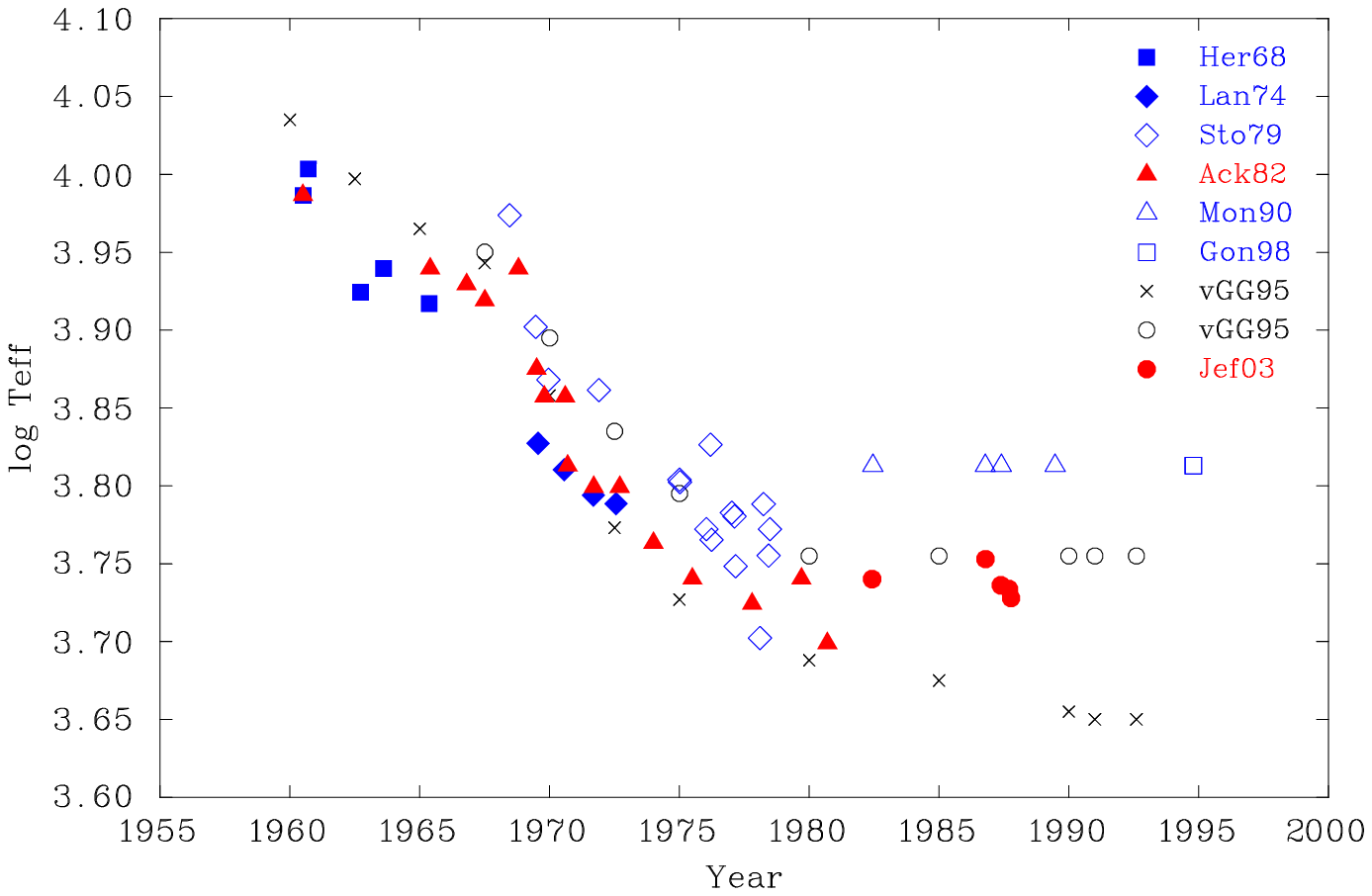,width=85mm} % ,bbllx=160,bblly=15,bburx=350,bbury=350}
 \caption[\object{FG\,Sge} \Teff]{The run of \te\ for \object{FG\,Sge}
  as determined spectroscopically and photometrically.
  Both low ($\times$) and high (dots) photometric \Teff\ scales 
  are from van Genderen \& Gautschy (\cite{vGG95}: vGG95).
  The spectroscopic measurements are from 
  Herbig \& Boyarchuk (\cite{Her68}: Her68),
  Langer et al.\ (\cite{Lan74}: Lan74),
  Stone (\cite{Sto79}: Sto79),
  Acker et al.\ (\cite{Ack82}: Ack82),
  Montesinos et al.\ (\cite{Mon90}: Mon90) and
  Gonzalez et al.\ (\cite{Gon98}: Gon98). New results from the
  present study are also shown (Jef03). This and
  other figures appear in colour in the online version of this paper;
  parenthetical references to colour refer to the latter.
 }
 \label{f:teff}
\end{figure}
%_____________________________________________________________________________

\section{The expansion of \object{FG\,Sge}}

The evolution track derived from optical photometry and 
pulsation analysis (van Genderen \& Gautschy \cite{vGG95}, Fig.\,\ref{f:evol})
is adopted as a principal
benchmark.  These authors actually give two evolution tracks, 
with low and high temperature scales respectively. We have 
adopted the high scale for reasons which will become clear later, 
although van Genderen \& Gautschy (1995) prefer the low scale. 
Underlying this is the basic assumption that 
over a long timescale and in the absence of major events affecting the
internal structure of the star, the evolution track will be 
smooth (Fig.~\ref{f:evol}). Superimposed on the evolution are short-term
pulsational changes ($\Delta V\sim0.15$, $\Delta (B-V)\sim0.04$), 
which will perturb instantaneous measurements
of \Teff\ and $L$. Other authors (Montesinos et al.\ \cite{Mon90},
Gonzalez et al.\ \cite{Gon98})
argue that the adopted $B-V$ calibration is unreliable 
because of line-blanketing in the optical by enhanced 
s-process elements. This does not apply to observations
before approximately 1970, when the first s-process episode was claimed 
(Langer et al.\ \cite{Lan74}). In this paper we investigate the
influence of s-process element enhancements on the $(B-V) - \Teff$ 
calibration.

It is also useful to check broad-band photometric measurements of \Teff\ 
against other methods. Spectrophotometry and spectroscopy  
have both been used to measure \Teff\ for \object{FG\,Sge} over the
last forty years.

Spectrophotometry of \object{FG\,Sge} covering the decade 1968--1978 
was calibrated against a sequence of MK standard supergiants (Stone \cite{Sto79}). 
Only wavelengths $>5\,556$\,\AA\ were used in an effort to avoid the effect of
blanketing by s-process elements. The effective temperatures inferred were 
$\sim 500\kelvin$ higher than those derived spectroscopically (Langer
et al.\ \cite{Lan74},
Fig.~\ref{f:teff}).  
The systematic difference could be due to non-LTE effects in the
atmosphere or, more likely in the present authors' view, to the influence 
of rare-earth metal lines on the red spectrum. These would have 
different consequences for the two 
methods of  temperature measurement. However the measured 
cooling rate  (310\kelvin\,yr$^{-1}$) agreed with
contemporary measurements (Smolinski, Climenhaga \& Kipper \cite{Smo76}). This rate
appeared to reduce later to $\sim170$\kelvin\,yr$^{-1}$
(Cohen \& Phillips \cite{Coh80}). 

We note that the spectroscopic measurements of effective
temperature \Teff\ do not
always agree well with measurements of the {\it ionization}
temperature $T_{\rm ion}$ measured from the same data 
(Langer et al.\ \cite{Lan74}) but are in satisfactory agreement with
independent contemporary analyses (Chalonge, Divan \& Mirzoyan \cite{Cha77}).

Spectral type alone can be a useful measure of \Teff\, if the
calibration (e.g.\ Johnson \cite{Joh66}) is reliable, and gives results
(Acker, Jaschek \& Gleizes \cite{Ack82})
that agree very well with the photometric calibration
(Fig.~\ref{f:teff}). It is important to note that the lowest
temperatures are given for the interval 1975--1981. 
The authors are not aware of any subsequent spectroscopic
measurement of \Teff\ that gives a value below 6\,000\kelvin.

The addition of ultraviolet 
and infrared photometry during the 1980's provided more
data with which to measure  \Teff\ using the total flux
method. However the intrinsic short-term variability
of \object{FG\,Sge} made simultaneous UV, visual and IR 
photometry difficult to achieve. The problem is most apparent in
the infrared, where a 115\,d intrinsic variation (Arkhipova \cite{Ark93}) 
was shown to be due to
a 1300\kelvin\ fluctuation in \Teff\ and a factor two change in
luminosity (Taranova \cite{Tar87}). 

The only study to exploit the IUE data
(Montesinos et al. \cite{Mon90}) used a simple comparison with standard model
atmospheres (Kurucz \cite{Kur79}) to provide estimates of \Teff, 
concluding that $\Teff=6\,500$~\kelvin\ and $\sg=1.5$ in the interval
1982--1989.  These authors also conclude that the previous
expansion had stopped sometime in the early 1980's and 
that a possible rise in temperature had occurred. The questions 
raised by this study concern the appropriateness of the model 
atmospheres used, the influence of variability on the comparison
of non-contemporary data and -- as was to become clearer in the 1990's
-- the development of a dust shell masking the photosphere. 

Subsequently, colours in 1991--1993 were reported to be 
consistent with spectral type Gp, while the line
spectrum  remained essentially the
same as in 1983 (Stone, Kraft \& Prossner \cite{Sto93}). 1992 August marked the
beginning of a new phase in the evolution of \object{FG\,Sge}
(Papousek \cite{Pap92}) 
in which it has shown a series of deep obscurational minima 
reminiscent of an R\,CrB-type light curve (Royer \cite{Roy99}).
The clouds of dust formed during these events have substantially
transformed every subsequent observation of the underlying star,
so that it has been difficult to follow any further evolution in,
for example,  \Teff\ or intrinsic brightness. 

Several authors have employed the intrinsic variability 
or pulsations  of \object{FG\,Sge} as a proxy  for \Teff.
The principle is that linear pulsation models provide a
relation between mass, period, luminosity and temperature.
Assuming that the star evolves at constant luminosity (and mass), 
then a pulsation period gives \Teff. However, the difficulty of
establishing reliable periods and the necessary assumptions 
regarding mass and luminosity suggest these estimates should
be used with caution.  

\section{The chemical evolution of \object{FG\,Sge}}

The expansion of FG\,Sge has naturally caused huge changes in its
spectrum. It has been argued that these are not purely due to
changes in \Teff\ alone, but that there have been concomitant changes
in the chemical composition of the photosphere. The most significant
studies are reviewed here. It is appropriate
to note that each study considers a different epoch, using a different
observational dataset. Of equal note, each study approaches the
abundance analysis using a different method with different underlying
assumptions. The dangers of simply concatenating such a series of 
results must be self-evident. 
Attention is drawn to some  of the major inconsistencies.

\subsection{Herbig \& Boyarchuk \cite{Her68}}

The first data to permit a view of the chemical evolution of
\object{FG\,Sge} was provided by Coud\'e spectrograms obtained 
at the Lick Observatory between 1960 and 1972 (Herbig \& Boyarchuk \cite{Her68}; 
Langer et al.\ \cite{Lan74}), together with some earlier low-dispersion plates from
1955. The dramatic spectral evolution from B4\,I to A5\,Ia was
initially interpreted as an expanding surface, or false photosphere, 
above an invisible central star (Herbig \& Boyarchuk \cite{Her68}). However, 
it is the extensive
tables of equivalent widths published by these authors
that provide the most valuable and accessible data from which to 
reassess the history of \object{FG\,Sge}. These include
data from absorption lines between 3300 and 4930 \AA, including those
due to hydrogen and neutral helium, light elements and iron-group ions. 

These equivalent widths were used by Herbig \& Boyarchuk (\cite{Her68}) in a 
curve-of-growth analysis to measure abundances of 20 species from
hydrogen up to europium. The results were tabulated relative to
\object{$\alpha$\,Cyg}, after normalising all abundances so that 
the mean combined abundance of Si, Ti, Cr and Fe was the same for both
FG\,Sge and \object{$\alpha$\,Cyg} (see appendix \ref{a:hb68}). 
The following points are interesting:\\
i) The absence of \ion{He}{i} lines from 1962 onwards places limits on
\Teff\ and helium abundance;\\
ii) a probable excess of carbon (+0.7 dex) in most spectra;\\
iii) a 0.5 dex overabundance of oxygen;\\
iv) an apparent excess of cobalt and strontium (+0.8 dex) at
   all epochs;\\
v) nearly all other elements are overabundant by 0.2 to 0.3 dex.  

Assuming that the differential abundance measurements are relatively
secure, a modern abundance analysis of
\object{$\alpha$\,Cyg} can be used (e.g.\ Albayrak \cite{Alb00}) to compute
absolute abundances (see Appendix \ref{a:hb68}). 
After correcting for substantial differences between 
\object{$\alpha$\,Cyg} and the Sun, we could further conclude:\\
i) carbon is overabundant by 1 dex, \\
ii) helium, nitrogen -- iron and nickel are near-normal, and \\
iii) cobalt, strontium, zirconium, barium and europium are 
overabundant by from +0.7 (cobalt) to +2.1 dex (strontium). 

It is significant that large overabundances in heavy elements,
particularly cobalt and strontium, are evident from 1960 onwards. 
We should, however, be cautious in evaluating these data:
measurement errors were not given, and the curve-of-growth method 
has limitations. However, the equivalent widths themselves are
invaluable as they permit further analysis using modern
model-atmosphere techniques.

\subsection{Langer, Kraft \& Anderson \cite{Lan74}}

As the star continued to evolve, it began to show abnormally
strong lines of \ion{Y}{ii}, \ion{Zr}{ii}, \ion{Ce}{ii}, \ion{La}{ii}
and other s-process species sometime after 1967 (Langer et al.\ \cite{Lan74}).
Apparent abundances rose to some 25 times (+1.4 dex) the solar value.
Commencing with a reanalysis of the final
spectrum used by Herbig \& Boyarchuk (\cite{Her68}),  the most interesting 
results are:\\
i) a drop to near-normal carbon abundance (low weight). \\
ii) steady Ti, Cr and Fe abundances. \\
iii) significant increases in Y and Zr. \\
iv) the appearance of La, Ce, Pr, Nd and Sm with +0.9 \dots +1.5 
dex overabundances.

It is difficult, however, to reproduce these values as the crucial
equivalent widths have not been published, and only a photographic
reproduction of the coud\'e spectra in a small region of the spectrum
is accessible to us\footnote{The authors have contacted two 
coauthors of Langer et al.\ (Kraft and Anderson) and the principal author
of Cohen \& Phillips (\cite{Coh80}), who also had access
to these data. Detailed records of the original work can no longer be located. 
Although the original
plates should still be in the Lick plate archive, their extraction and
analysis lie beyond our current resources. }. 
Taken together, these two papers (Herbig \& Boyarchuk \cite{Her68}; Langer et al.\ \cite{Lan74}) pose a
number of significant questions:

\noindent i) What has happened to the carbon abundance in FG\,Sge? It
appeared to start high and then fall. However, if dredge-up of
s-process elements occurred {\it during} this epoch, then carbon      
should also have been {\it enriched}. This may not be a severe
observational constraint; the early carbon abundances were {\it ``based
on one or two very high excitation {\rm C}{\sc i} lines in an uncertain
atmosphere with $kT\sim0.5$ {\rm eV}, so the Boltzmann factor would be pretty
sizeable, and the results pretty uncertain''} (Kraft, private communication).
Since the carbon abundance derived by Langer et al.\ (1974) is also quite
uncertain, the case of the carbon abundance as based on these two 
investigations remains unsettled.

\noindent ii) Is FG\,Sge hydrogen-deficient and when did it become so?
Is there any corollary with, say, V4334\,Sgr, claimed to show a 
similar transition to hydrogen-deficiency during its redward evolution
(Asplund et al. \cite{Asp97})? Deep envelope mixing
and the enrichment of s-process material should be dominated by the
dredge-up of helium and  dilution of surface hydrogen. This is a very
important question. A low hydrogen abundance would be difficult to 
detect after 1960, yet it would have a dramatic effect on the
structure of the atmosphere and the strengths of many lines in
the spectrum. 

\noindent iii) Has FG\,Sge been s-process rich all the time? 
Are early enhancements in Co and Sr 
symptomatic of a general s-process enhancement that was simply not
recognised until the photosphere cooled sufficiently for ions with lower 
ionization potentials to become visible?  

\noindent iv) How would the adopted model atmosphere affect, 
for example, measurements of the change of abundance with effective 
temperature? In
the domain of interest, a very small drop in \Teff\ produces a
very sharp increase in \Wl\ for the second ions of rare-earth
elements. Would additional opacity from carbon, or reduced opacity
from hydrogen depletion be sufficient to cause a significant change 
in the original conclusions?
 
\subsection{Cohen \& Phillips \cite{Coh80}}

Cohen \& Phillips (\cite{Coh80}) continued the work of Langer et al.\ (\cite{Lan74}) 
using 4-m echelle spectra obtained each year from 1975 to 1978. They
introduced a ratio technique to deduce \Teff\ and chemical abundances
from each spectrum. No lines with $\Wl>200$~m\AA\ changed 
by more than 40\% during this epoch. They also found for strong lines
that $\Wl^{ij}\propto(N_{ij})^{\beta}$, with $\beta=0.30$ rather than
the anticipated 0.50. Whether this is a 
consequence of the low surface gravity or
of the composition of the photosphere remains to be verified. The
authors' inference from the upper limit to changes in \Wl\ was that, 
in the absence of any cooling, the abundances of
all elements with strong lines, including the rare earths, did not 
increase by more than 0.50 dex over the 3 yr time span. Any cooling
would reduce this maximum abundance change. One comment is 
notable; {\it ``These models \ldots\ predict for solar abundances of the
non-$s$-process 
elements (believed to be valid for at least the even iron-peak
elements
in \object{FG\,Sge}) values of \Wl\ smaller than those observed in the
spectra.''} The authors suggest that this is due to the extended nature of
the atmosphere. It is also the classic symptom of an atmosphere with a
lower than usual continuum opacity as would be occasioned, for
instance, by a low hydrogen abundance. If it is not an abundance 
effect, the cause should have the same consequence for all lines, 
including the $s$-process elements, accounting for some, if not all, of
their apparent overabundance. Unfortunately, due to severe
line crowding in the blue, the spectra used in this study
were obtained longward of 5000\AA\, so that the only Balmer line
observed was H$\alpha$.

\subsection{Acker, Jaschek \& Gleizes \cite{Ack82}, Cowley, Jaschek
  \& Acker \cite{Cow85}}
 
Following the work at Lick Observatory, the spectrum of 
\object{FG\,Sge} during 1979 -- 1980 suggested a remarkable new 
phenomenon. In addition to the
increase in s-process elements (barium and the lanthanides), 
the abundances of iron-group
elements were significantly reduced (Acker et al.\ \cite{Ack82}; Cowley
et al.\ \cite{Cow85}).

\subsection{Kipper \& Kipper \cite{Kip93}, 
            Kipper \& Klochkova \cite{kip.klo01}}     \label{s:kip.kip}

As well as abundance estimates for magnesium, four iron-peak elements
and 13 rare-earth elements, Kipper \& Kipper (\cite{Kip93}) report the
detection of C$_2$ in 1992. The deduced abundance of carbon is dependent 
entirely on the assumed \Teff\ (and presumably on the assumed hydrogen
abundance) but appears to be enhanced, lying in the range $ -2.70
(5\,500 \kelvin) < \log N({\rm C}) < -1.6 (6\,500 \kelvin)$, where
$N({\rm C})$ represents the fractional abundance (by number of atoms) 
  relative to the total of all elements.
  Consequently,
carbon is considered to have increased along with or after the $s$-process 
elements. These authors used $\log g=1.0$ in their analysis.

High-resolution spectra taken in 2000 and analysed with the same 
      model atmosphere ($5\,500\kelvin / \log g=1.0$) gave essentially the 
      same results (Kipper \& Klochkova \cite{kip.klo01}).  The oxygen 
      abundance, based on one single [O\,{\sc i}] line, resulted in 
      $N({\rm C})/N({\rm O})=2.5$. 
      Note that the model parameters imply $\log L/M \simeq 3.5$ 
      (solar units).

Although a hydrogen-deficiency 
      was not reported in either study, it should be noted that the abundances
      deduced are likely to be substantially smaller if hydrogen is
      significantly depleted. 

\subsection{Gonzalez et al. \cite{Gon98}}
\label{s:gonz.etal}

\begin{figure}
\epsfig{file=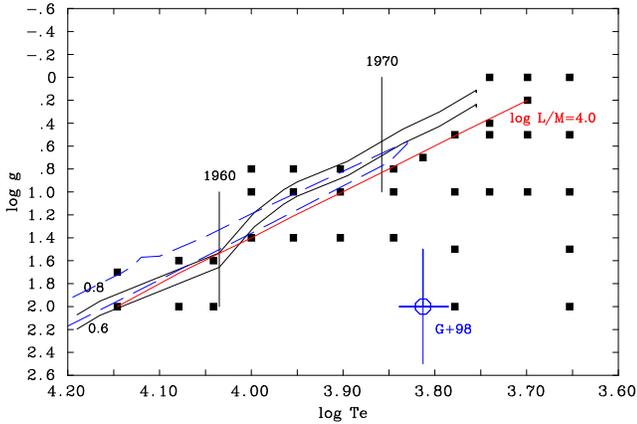,width=85mm} % ,bbllx=160,bblly=15,bburx=350,bbury=350}
 \caption[\object{FG\,Sge} evolution]{The run of \te\ and
  \sg\ for \object{FG\,Sge} (van Genderen \& Gautschy \cite{vGG95}; solid lines) 
  assuming $M=0.8\Msolar$ (below) and $0.6\Msolar$ (above). 
  Solid squares represent a selection of converged {\sc sterne}
  models, including the low-gravity limits. Several models with
  intermediate gravities were also computed. 
  The straight line (online: red) indicates a 
  sequence of models with $\log L/M=4.0$,  for which 
  theoretical colour indices, emergent spectra and equivalent widths 
  were computed. The dashed line (online: blue) shows part of the redward loop
  from the evolutionary
  track of a $0.625\Msolar$ post-AGB model suffering from a late
  thermal pulse (see section \ref{s:ltp}). The circle (online: blue)
  shows the model adopted by Gonzalez et al. (\cite{Gon98}) for their 1994 October
  21 spectrum (Section \ref{s:gonz.etal}).
 }
 \label{f:grid}
\end{figure}

Gonzalez et al.\ (\cite{Gon98}) obtained modern high-resolution
high S/N spectra between 1992 and 1996. Their fine
analysis considered primarily a spectrum obtained in 1994.  
From \ion{Fe}{i} and \ion{Fe}{ii} lines they deduced
$\Teff=6\,500\pm400\kelvin$, $\sg=2.0\pm0.5$, 
  (Fig.~\ref{f:grid})
and a 3 dex
enhancement of s-process elements. About 1~dex of this enhancement must 
have occurred between 1992 and 1994 (Kipper \& Kipper \cite{Kip93}). 

As with the photometric
analysis of Montesinos et al.\ (\cite{Mon90}),  the difficulty posed 
by this analysis
is that it requires the star to have both heated -- possibly 
as it returns towards higher temperatures -- and -- to yield the
measured surface gravity --  shrunk in radius
by a factor of $\sim10$ since the observations reported by Acker et
al.\ (\cite{Ack82}). Such a change would have produced a $\sim1.7$ dex
  reduction in luminosity, but no concomitant change in $V$ (4
  magnitudes) was observed over this
  interval. 

It is therefore puzzling that emphasis should have been given
to the 3 dex abundance change, but not to the radius change. 
The effective temperature and abundance measurements are not
unaffected by this contradiction, since both depend on the surface 
gravity and model atmosphere adopted. 
Reviewing our own work on atmospheric fine analyses we find that, in
general, lower surface gravities give lower effective temperatures
for the same ionization or excitation equilibrium with $d \log T_{\rm eff} / d
\log g \sim 0.06$. Reducing the Gonzalez et al. (\cite{Gon98}) surface
gravity by the requisite amount to give $\log L/M \approx 4.0$ yields
$T_{\rm eff} \sim 5\,500$K. 

Secondly, if the abundances are as far from normal as these authors 
suggest, the implications of reduced hydrogen opacity and increased 
metal opacity will be extreme, as Gonzalez et al. (\cite{Gon98}) were aware. 
The structure of the adopted ATLAS9 
model atmospheres with normal hydrogen and metal abundances would have 
been quite inappropriate for modelling the spectrum of such a peculiar
star. With few reliable temperature indicators available to Gonzalez
et al. (\cite{Gon98}), 
one wonders if the starting premise that the  
Montesinos et al. (\cite{Mon90}) results were
{\it "probably the most accurate published temperature estimates since 
the early 1970's"}  was unfortunate.

However, even if both effective temperature and surface gravity are
wrong, the abundance measurements may not be so badly in error. 
If they are not, then FG\,Sge will have taught us something 
important about the late stages of stellar evolution. 
It is therefore both legitimate and necessary to make 
a thorough reappraisal of the chemical evolution of FG\,Sge, including 
those measurements made with the most modern equipment and methods.

%______________________________________________________________________________
% Figure "grid"   Evolution track and model grid
%
%_____________________________________________________________________________

\section{Model atmospheres and synthetic spectra}

In order to interpret the evolving spectrum of \object{FG\,Sge} it is necessary
to simulate the emergent radiation observed through its history
under a range of assumptions. The principal constraint
on these is the evolutionary track derived from optical colours
and pulsation periods (van Genderen \& Gautschy \cite{vGG95}). This track is shown in 
Fig.~\ref{f:grid}, assuming masses of 0.6 and 0.8 \Msolar, respectively. 

It will be seen that these tracks lie
close to the classical Eddington limit and that obtaining suitable model
atmospheres is not trivial. In the current era, the strong radiation 
field and extended atmospheres generally encountered in A-type supergiants
would be considered to require a treatment of the failure of
local thermodynamic equilibrium (LTE) and the curvature of the stellar 
photosphere, in addition to a luminosity-related stellar wind.  
However, in the present case, the quality of the available
data are insufficient to warrant such effort so that the adoption of
line-blanketed plane parallel model atmospheres in LTE may be
considered a satisfactory approximation.

The program {\sc sterne} (Sch\"onberner \& Wolf \cite{Sch74}; Jeffery,
Woolf \& Pollacco \cite{Jef01})
can compute low-gravity model atmospheres over a wide range of
\Teff\ and chemical composition. Its use provides control over the
assumed physics and allows us to monitor the convergence
of each model atmosphere. We note that the  
current opacity distribution functions are incomplete,
being based on the Kurucz \& Peytremann (\cite{Kur75}) line list, 
and that the contribution of molecules to the equation of state and opacity
in low \Teff\ atmospheres is not considered. Since the calculations
we make are largely comparative, the fact that they are used 
consistently to discuss all of the data for FG\,Sge across a large
range in \Teff\ far outweighs these limitations. 

Model atmospheres were computed assuming a standard (solar) mix of 
elements heavier than helium. Three grids were computed assuming
number abundances of hydrogen and helium $(\nH, \nHe) = (0.9,0.10), 
(0.10, 0.90)$ and (0.01,0.99), which we have labelled hydrogen-normal, 
hydrogen-poor and hydrogen-deficient, respectively.  The goal was to 
obtain a sequence of models with $\log (L/\Lsolar)/(M/\Msolar) \equiv
\log L/M = 4.0$, close to the observed value and  
spanning the range in \Teff\ exhibited by \object{FG\,Sge} since 1960.  
The grid computed (although not every model) is illustrated in terms of 
\Teff\ and \sg\ in Fig.~\ref{f:grid}.
We note that FG\,Sge's luminosity-to-mass ratio may even be larger 
     than the one assumed here.

Within each of these model grids, the composition of the heavy
  element component assumed in the calculation of the equation 
  of state and the continuous (bound-free) opacity was (by number
  fraction relative to the total) : C:0.000363, N:0.000112, 
  O:0.000832, Mg:0.0000339, Al:0.00000269, Si:0.0000316, S:0.0000158, 
  Ca:0.00000229, and Fe:0.0000288, these being the standard solar 
  abundances of these elements at the time they were adopted into  
  {\sc sterne}. The abundance of the most abundant
  element (H or He) was reduced by the total of these to ensure a total number
  density of unity.  A complete review of the opacity sources used in {\sc
  sterne} has been given recently by Behara \& Jeffery (\cite{Beh06}). 
  The abundances used in the calculation of the 
  line-opacity were those hard-wired into the opacity distribution 
  functions (ODFs) - being the Kurucz (\cite{Kur70}) ATLAS6 ``p00'' 
  ODF for the hydrogen-rich grids
  and the M\"oller (\cite{Moe90}) ``he90'' ODF for the helium-rich
  grids. 

Attempts to compute corresponding models using  
Kurucz' program {\sc atlas9} (Kurucz \cite{Kur91}) were not so successful.
These generally failed at very low gravities due to negative pressures in 
the equation of state, although we note that Lamers \& Fitzpatrick
(\cite{Lam88}) 
had  computed much lower gravity models using an older version of 
{\sc atlas} (Kurucz \cite{Kur79}), achieving $\sg=1.2$ rather than 2.0 
at $\te=10\,000\kelvin$. 

Using model atmospheres from {\sc sterne}, the formal solution 
program {\sc spectrum} (Jeffery et al.\ \cite{Jef01}) permits the
computation of synthetic spectra (in LTE) over large wavelength intervals,
the computation of individual line profiles and curves of growth
and the computation of elemental line abundances for given equivalent 
widths. 

A database for transitions of light elements in hot stars is maintained by 
one of us ({\sc LTE\_LINES}, Jeffery et al.\ \cite{Jef01}). For the present
study, substantially more extensive data for iron-group and heavier
atoms  are required. 
The iron-group data have been collated from the extensive line lists of Kurucz
(\cite{KurCD01}), providing a total of over 144\,500 metallic
absorption lines with wavelengths between 2900 and 8000 \AA. 
For s-process elements, data for over 11\,500 lines have been recovered
from  the Vienna Atomic Line Database (Piskunov et al.\ \cite{VALD1};
Kupka et al.\ \cite{VALD2}). Since this represents only a subset of 
the s-process element lines observed in FG\,Sge, the effects of these elements will
be underestimated.

%______________________________________________________________________________
% Figure "colours"   Comparison of H-normal and H-poor atmospheres
%
\begin{figure}
\epsfig{file=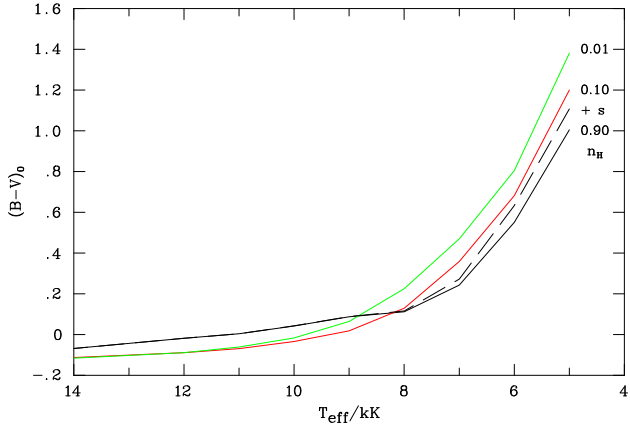,width=85mm} % ,bbllx=160,bblly=15,bburx=350,bbury=350}
 \caption[Model colours]{Simulated colour indices ($B-V$) for a
 sequence  of model atmospheres with $\log L/M=4$ and  
 normal hydrogen (solid: black), 
 10\% hydrogen (by number, solid: red), 1\% hydrogen (solid: green)
 and normal hydrogen with enhanced s-process elements 
 (broken: black).  }
 \label{f:cols}
\end{figure}
%_____________________________________________________________________________

\section{Key questions}

This section starts to address some of the key questions raised in
the previous discussion. For example: \begin{itemize}
\item Why are there such substantial discrepancies in the \Teff\ scales? 
\item What was the evolution of the hydrogen abundance? 
\item Is there any evidence of carbon enrichment ? 
\item Did s-process enhancement ever really occur? 
\end{itemize}

\subsection{Are colour indices sensitive to composition?}

The {\sc sterne} model atmospheres include a 
calculation of the emergent spectrum sampled at 342 
wavelengths from the X-ray to the infrared, with sampling
approximately every 20~\AA\ in the UV and visible.  A consequence 
immediately apparent is that for $\Teff<10\,000\kelvin$, the hydrogen-deficient
models are substantially redder, in the optical region, than the
hydrogen-normal models.

These flux distributions have been augmented by computing 
detailed synthetic spectra with {\sc spectrum} using the linelists
described above and 200,000 frequency points between 2900 and
  8000 \AA. These were convolved with 
appropriate detector response
and filter transmission functions to compute theoretical 
intrinsic colour indices $(U-B)_0$ and $(B-V)_0$.
for models with both normal and modified abundances. 
In addition to the variations in hydrogen abundances, we also 
computed colours for a models in which the abundances of 
s-process elements 
strontium, yttrium, zirconium, barium, lanthanum, cerium,
praseodymium, neodymium. samarium, europium, terbium, dysprosium,
erbium, ytterbium, lutetium, hafnium, and lead 
were enhanced by two dex. Most of these
have been reported at some time to be strongly overabundant in
\object{FG\,Sge}.  

The results
for $(B-V)$ are shown in Fig.~\ref{f:cols} which confirms that the 
cooler hydrogen-deficient models may be {\it redder} at a
given temperature than their hydrogen-rich counterparts. This
is a direct consequence of the much lower continuum opacities in the
hydrogen-poor models. Strong metal
line blanketing due to singly-ionized iron-group elements in the
near-UV coincides with  the peak of the Planck function and has
a much stronger effect than in the hydrogen-rich model atmospheres -- the
{\it ratio} of line to continuous opacity is the crucial quantity.
The increase in metal-line blanketing is partially compounded in the 
hotter models by a reduction in the contribution from the high-order
Balmer series and Balmer continuum. One implication is that, for 
observations of
\object{FG\,Sge}, the assumption of a normal hydrogen abundance could
lead to an underestimate of \Teff\ by as much as $1\,000\kelvin$ 
for $\Teff\simle9\,000\kelvin$. 

The original object of this experiment was to check the effect of
enhanced s-process element abundances on the colour--\Teff\
calibration.  Although only a subset of all s-process lines was 
included,  Fig.~\ref{f:cols} shows that these effects are probably
not severe, contrary to the assertion by Montesinos et al. \cite{Mon90}.
They produce much smaller changes in $(B-V)$ than caused by changes 
in the hydrogen abundance. Therefore, if the hydrogen abundance is
normal, the existing $\Teff-(B-V)$ calibration (van Genderen \&
Gautschy \cite{vGG95}) should 
be acceptable.

%______________________________________________________________________________
% Figure "2color"   Comparison of H-normal and H-poor atmospheres
%
\begin{figure}
\epsfig{file=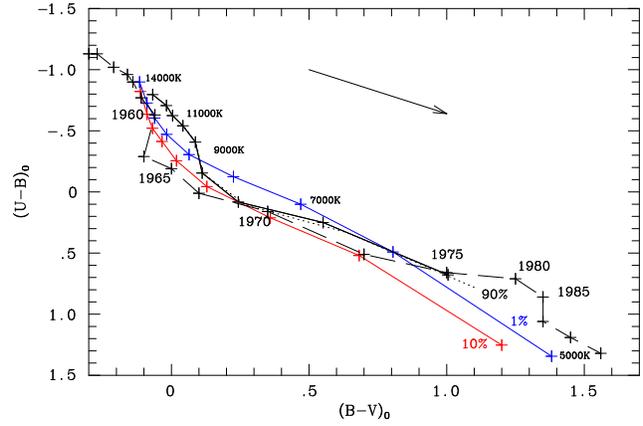,width=85mm} % ,bbllx=160,bblly=15,bburx=350,bbury=350}
 \caption[Model 2 colour diagram]{Simulated colour indices $(U-B)$ vs.\
 $(B-V)$ for a sequence of  model atmospheres with $\log L/M=4.0$: 
 1) normal hydrogen (black), 
 2) 10\% hydrogen (by number, red),
 3) 1\% hydrogen (blue) and
 4) normal hydrogen with s-process elements increased by 2 dex (black 
 dotted). 
 The hydrogen number fractions are marked at the cool end of each
 sequence. The effective temperatures of selected models are labelled 
 on the 1\% hydrogen sequence, models with the same \Teff\ are
 marked in each sequence. 
 The  evolution of \object{FG\,Sge} from 1960 to 1985 is also shown 
 (van Genderen \& Gautschy \cite{vGG95}, broken black). The 
 standard galactic reddening is shown as an arrow. }
 \label{f:2cols}
\end{figure}
%_____________________________________________________________________________

\subsection{Can photometry solve the hydrogen problem?} 

Taking the previous discussion one step further, it is
interesting to ask whether photometry alone can constrain 
the chemical evolution of  \object{FG\,Sge}. It has been shown
that the \Teff\ scale depends on the actual hydrogen
abundance. For early-type stars, the Balmer continuum dominates the 
position of a star in the two-colour $(U-B) - (B-V)$ diagram.
Figure~\ref{f:2cols} compares the locus of hydrogen-rich, poor and
deficient model sequences with the evolutionary track of \object{FG\,Sge}.  

While reducing the hydrogen from near-normal to
$\sim10\%$  reddens the evolutionary track in $(B-V)$,
as discussed above, the concomitant reduction in the Balmer continuum 
makes the tracks in the two-colour diagram distinguishable only in
selected temperature and abundance domains.

The only clear conclusion
is that, assuming a standard extinction law, \object{FG\,Sge} has
been too red for a normal composition since at least 1975. However,
since the star has been losing mass, its continued reddening may have
as much to do with nonstandard extinction as with changing
composition.

\subsection{How robust is the photometric calibration?}

If \Teff\ is inferred from $(B-V)$ alone, one has to compute $(B-V)_0$ 
assuming some value for the reddening \Ebv. Most authors have adopted
$\Ebv=0.4$, apparently following Herbig \& Boyarchuk (\cite{Her68}) who compared 
observed colours with those expected for the current spectral type. 
While these authors also reported an unusual value for
$\Eub/\Ebv=0.4$ rather than the more usual value between 0.7 and 0.8,
there have been no other suggestions of anomalous extinction 
before 1990, when dust-ejection episodes became frequent. It will be
seen that adopting $\Ebv=0.4$ is quite consistent with the optical/UV
flux distribution in the 1980's and discussed in Section \ref{s:iue.teff}.  

%______________________________________________________________________________
% Figure ``Herbig_EW0''  
%
\begin{figure*}
\epsfig{file=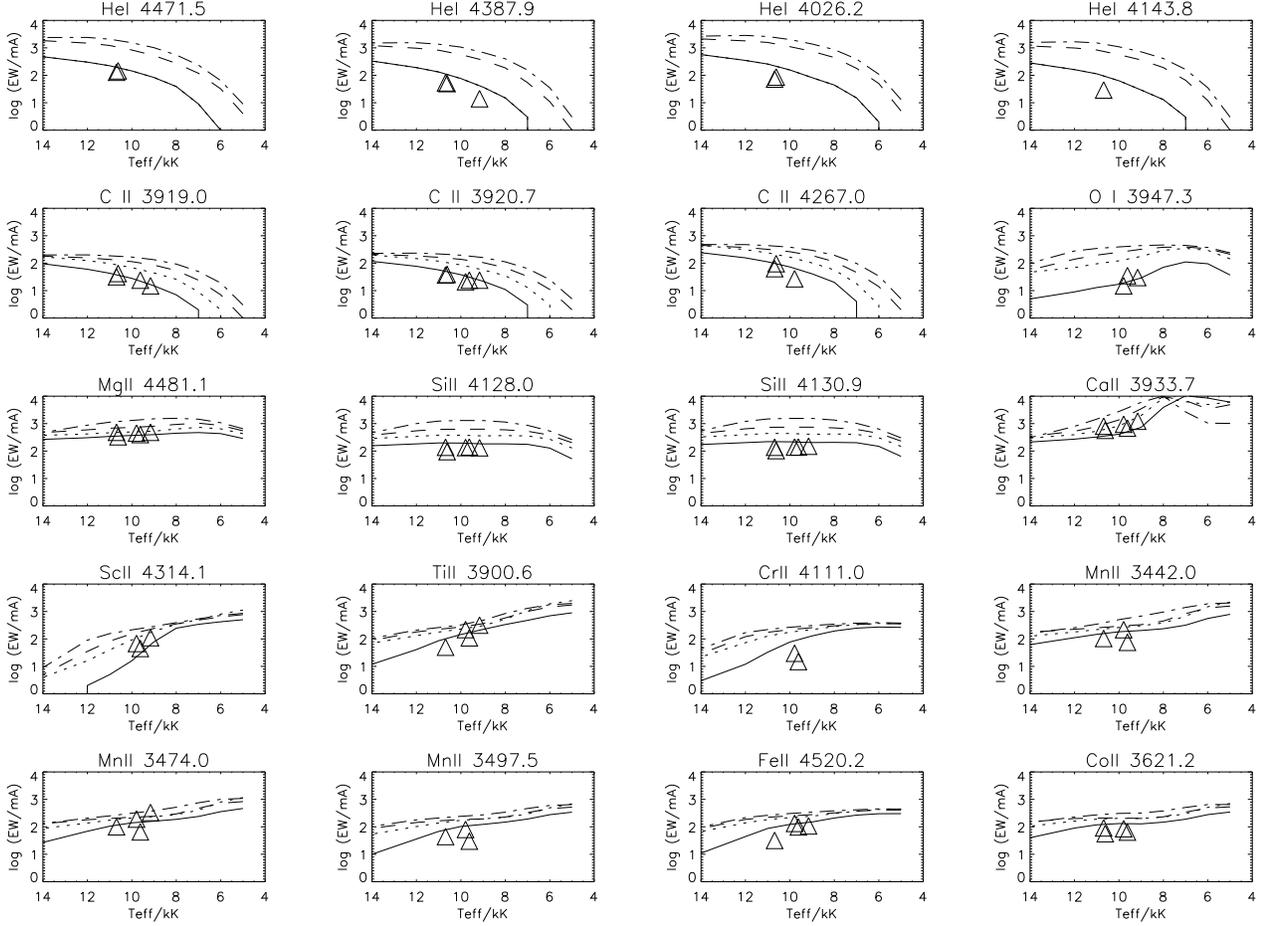,width=175mm} % ,bbllx=160,bblly=15,bburx=350,bbury=350}
 \caption[Herbig equivalent widths]{Equivalent widths of selected lines
  as a function of \Teff\ for model atmospheres with $\log L/M=4.0$ and
  $\nH=0.9$ (solid), 0.1 (dashed), and 0.01 (dot-dashed), assuming
  solar abundances and $\nHe=1-\nH$, and with $\nH=0.9$ and 
  $10\times\odot$ abundances (dotted, except \nHe). 
  Measured equivalent widths for \object{FG\,Sge} during 1960--1965
  are shown with symbols (Herbig \& Boyarchuk \cite{Her68}). }
 \label{f:hbew0}
\end{figure*}
%_____________________________________________________________________________
% Figure ``Herbig_EWS''  
%
\begin{figure*}
\epsfig{file=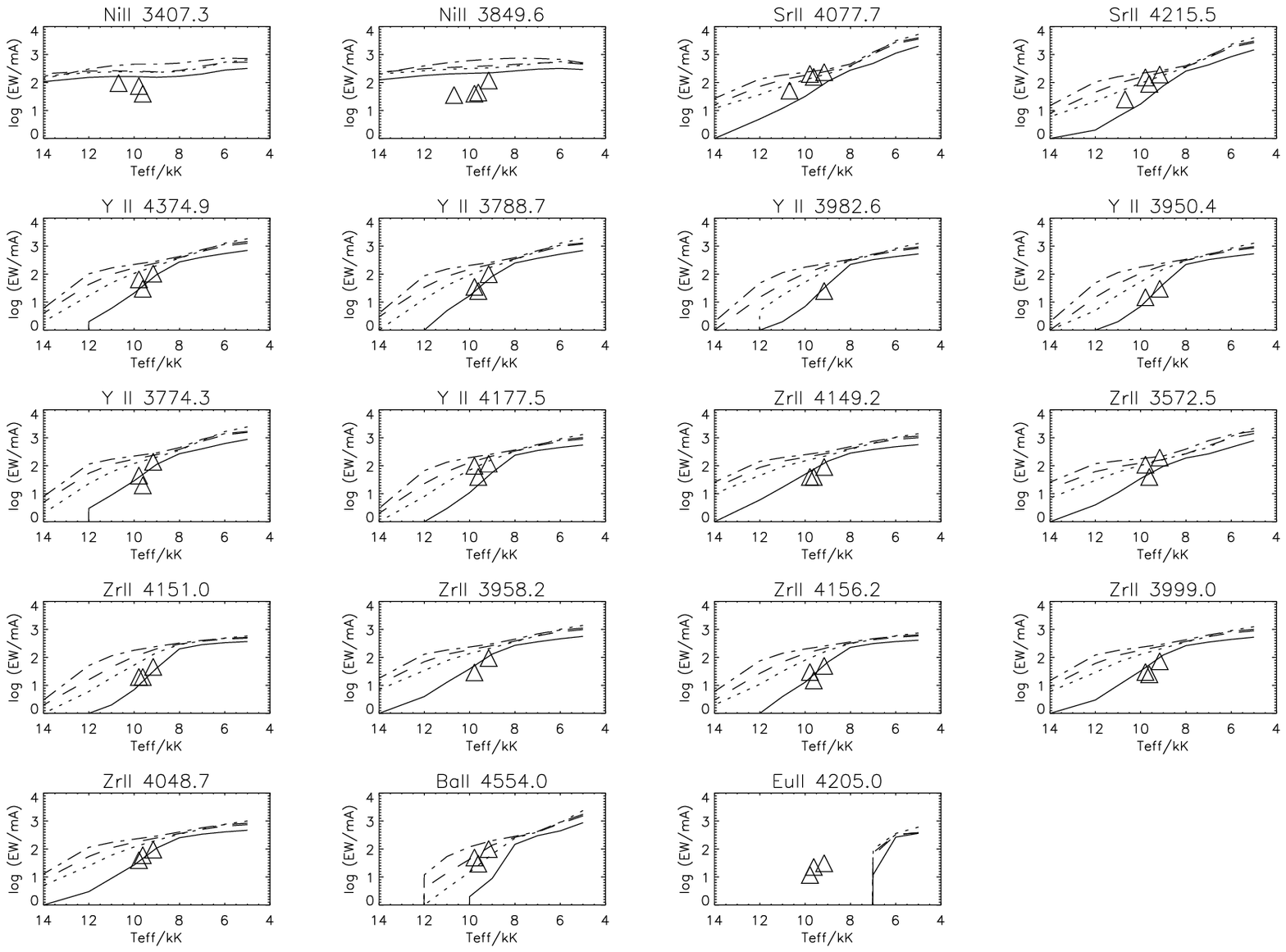,width=175mm} % ,bbllx=160,bblly=15,bburx=350,bbury=350}
 \caption[Herbig equivalent widths (S-process)]{As Fig.~\ref{f:hbew0}
 for lines of Ni and s-process elements. }
 \label{f:hbewS}
\end{figure*}
%_____________________________________________________________________________

\subsection{Was the composition of \object{FG\,Sge} unusual before 1965?}

The first step toward determining whether a substantial change in
surface composition actually occurred is to establish a zero
point. Specifically, was already
\object{FG\,Sge} depleted in hydrogen or rich in carbon or 
s-process elements in 1960--1965? 

Since Herbig \&  Boyarchuk (1968) provide equivalent width measurements for
over 300 lines and five different spectrograms, it is possible to
assess this question quantitatively. Rather than attempt
a detailed reanalysis of each spectrum, it is preferable to establish
what systematic effects can be established. 

We have computed predicted equivalent widths for 278 of the lines measured by 
Herbig \&  Boyarchuk (\cite{Her68}) for model atmospheres with $\log L/M=4.0$ 
in the range $14\,000\geq\Teff\geq5\,000$\kelvin\ and with hydrogen
compositions of 90\%, 10\% and 1\% by number. The abundances of all
species other than hydrogen and helium were assumed to be solar.
We also computed a set of equivalent widths for the 90\% hydrogen
models in which all species other than hydrogen and helium were
enhanced by a factor of ten. Microturbulent velocities $\vt=5$ and
10\kmsec\ were used. In Figs.~\ref{f:hbew0} and \ref{f:hbewS},
the equivalent widths for selected lines are plotted as a function of
temperature, with $\vt=5\kmsec$.  The observed equivalent widths for each 
line are also plotted, assuming \Teff\ given by the photometric calibration
(van Genderen \& Gautschy \cite{vGG95}, Fig.~\ref{f:evol}). The maximum uncertainty in \Teff\ 
is $\pm500$\kelvin.
In particular, these figures demonstrate how sensitive different
lines are to changes in \Teff, hydrogen abundance and the elemental abundance.

A recurring question through the following analyses is that of the
microturbulent velocity. The appropriate value to choose for this 
artificial factor is related to \Teff\ and \sg\ (Gray, Graham \& Hoyt \cite{Gray01}), 
and also to the specific ion in question (Albayrak \cite{Alb00}). To 
constrain \vt\ in a problem where the number of degrees of freedom
is comparable with the quantity of available data, we have
chosen \vt\ to be either 5 or 10 \kmsec\ according to which gives 
the most conservative interpretation. In practice, this 
meant using $\vt=5\kmsec$ when \object{FG\,Sge} was an early A star, 
and  $\vt=10\kmsec$ subsequently.

A number of strong conclusions concerning
the following elements in  \object{FG\,Sge} in 1960--1965 may be
drawn:
\begin{itemize} 
\item He is  normal. 
\item C and O are not detectably different from normal, but 
      C/O $> 1$ cannot be excluded.
\item Sr is significantly above normal ($\sim\times10$).
\item Y and Zr are not significantly different from normal.
\end{itemize}
Weaker conclusions include the following:
\begin{itemize} 
\item Mg, Si, and Ca are not well determined at these temperatures.
\item Sc and Ti appear normal.
\item Mn, Fe, Co and Ni may be somewhat underabundant.
\item Ba and Eu are significantly above normal ($\sim\times10$, one
absorption line only for each species.
\end{itemize}
These conclusions are in line with the differential measurements given
by Herbig \& Boyarchuk (\cite{Her68}) if the abundance differences 
between $\alpha$ Cyg and the Sun are also taken into account. 

%______________________________________________________________________________
% Figure "langer"   Comparison of Langer simulations
%
\begin{figure*}
\epsfig{file=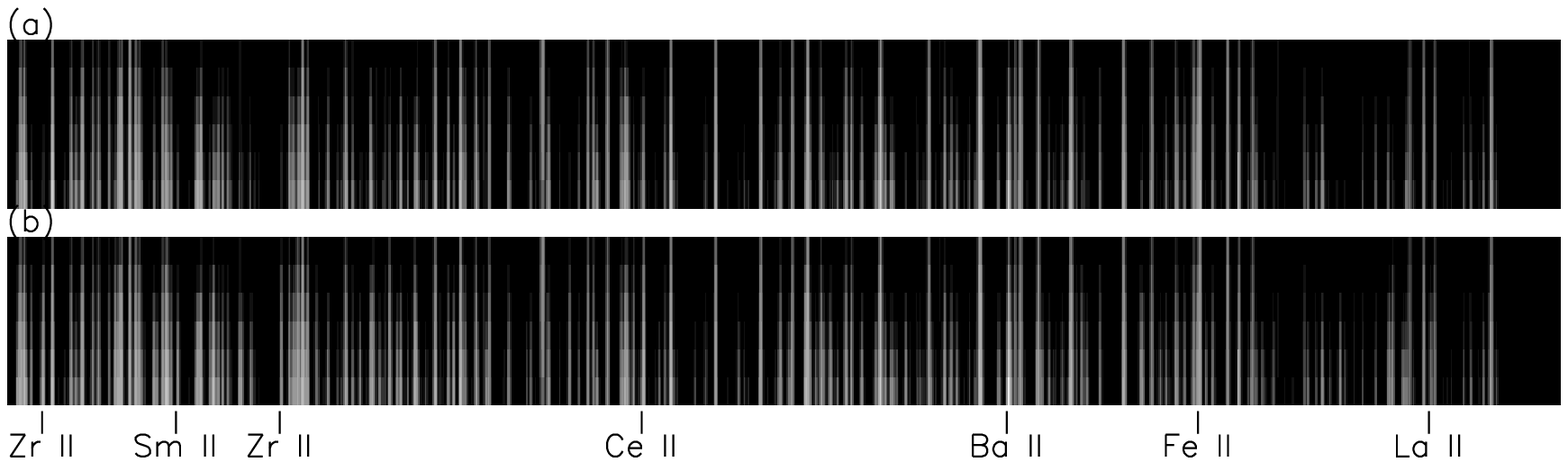,width=180mm} %,bbllx=0,bblly=735,bburx=240,bbury=60}
 \caption[Langer spectrogram]{Simulations of the blue
 region (4398--4640\,\AA) of the spectrum of FG\,Sge between 1965 and
 1972 assuming  
 (a) solar abundances and
 (b) abundances from Langer et al.\ (\cite{Lan74}).
 Each panel shows simulations
 with \Teff\ decreasing from 8\,000\,K (Sp A8I: top) to 5\,500\,K 
 (Sp G0I: bottom) in steps of $500\,{\rm K}$, and has
 been plotted for comparison with Fig.\,1 of Langer et
 al.\ (\cite{Lan74}). The models have been degraded to an instrumental resolution of 
 0.25\,\AA.  Note the progressive enhancement with reduced \Teff\ of lines due
 to s-process elements compared with lines of iron-peak elements. }
 \label{f:langer}
\end{figure*}
%______________________________________________________________________________
\begin{table}
\caption{Elemental abundances used in Fig.~\ref{f:langer}, written as
         $ \log N_{\rm el} - \log N_{\rm H} + 12 $, compared with
         solar photospheric abundances by Asplund, Grevesse \&
         Sauval (\cite{Asp05}).}
\label{t:langer}
\begin{center}
\begin{tabular}{ c r r }
\hline\hline\noalign{\smallskip}
Elem &  FG\,Sge  & Sun  \\
\noalign{\smallskip}\hline\noalign{\smallskip}
Fe   &  7.25 & 7.45 \\
Sr   &  3.78 & 2.92 \\
 Y   &  3.56 & 2.21 \\
Zr   &  3.88 & 2.59 \\
Ba   &  2.80 & 2.17 \\
La   &  2.69 & 1.13 \\
Ce   &  2.85 & 1.58 \\
Sm   &  2.00 & 1.01 \\
\noalign{\smallskip}\hline
\end{tabular}
\end{center}
\end{table}

\subsection{Was there an s-process enhancement episode around 1965--1970?}

If some s-process elements were already enhanced in the interval 
1960--1965, what is the significance of the dramatic increase
reported between 1965 and 1970 (Langer et al.\ \cite{Lan74})? The evidence rests on
a reanalysis of the plate obtained in 1965 by Herbig \& Boyarchuk (1968)
in which both light and heavy s-process elements are reported to be low. It is
appropriate to note the remark by Langer et al.\ that 
{\it ``In a first reconnaissance
of the abundance changes \ldots, rather crude methods seem entirely 
appropriate \ldots''}.
The 1965 spectrum only contains one line of barium amongst the
heavy elements considered to be normal in 1965 by Langer et al.

We have already found this line and one due to europium to 
be stronger than expected. 
 Whilst we find that yttrium and 
zirconium are not detectably different from normal in 1965, we do find another 
light s-process element, strontium, to be about 1\,dex overabundant in 1965, 
before the apparent enhancement episode began.

We have considered the question from two angles. First, 
we have calculated theoretical equivalent widths of several lines
for the given s-process abundances and \Teff. By deducing 
what \Teff\ or \vt\ would be required if the abundances were to be
normal and yield the same equivalent widths, 
we conclude that it would have been be impossible to confuse 
an s-process overabundance with a systematic error in \Teff\ or a
significant difference in \vt\ between FG\,Sge and the comparison 
stars $\alpha$\,Per and $\alpha$\,Cyg.  

Second, we have attempted to reconstruct the 
photographic spectra presented by Langer et al.\ (\cite{Lan74}) in
Fig.\,\ref{f:langer}. The logarithmic greyscale images represent 
the region 4398 to 4640 \AA, degraded to an instrumental resolution of
0.25\,\AA\, judged to correspond approximately to 8 \AA\,mm$^{-1}$
dispersion plates.
Spectra were computed for \Teff\ from 8\,000 K to 5\,500 K in steps of
500 K,  for two chemical mixtures, one with solar abundances  and 
one with abundances reported for 1972 August (Langer et al. \cite{Lan74}, Table
\ref{t:langer}), 
and for $\vt=5$ and $10\kmsec$. Those with  $\vt=10\kmsec$ are shown in
Fig.\,\ref{f:langer}. The
temperature sequence includes spectral classes from A8\,I to G0\,I, 
bracketing the Langer et al. range  F0\,I to F6\,I. 
The progressive enhancement with falling \Teff\ of lines due
 to s-process elements compared with lines of iron-peak elements is
evident, and confirms that s-process elements were overabundant. 

From the limited data available, we draw the following conclusions
for the interval 1965--1972.
\begin{itemize}
\item the Langer et al.\ results are broadly reliable.
\item abundances of light and iron-group elements are near normal.
\item Y and Zr are $\sim1$ dex overabundant and had apparently increased
  after 1965. Note that Sr had been $\sim1$ dex overabundant
  before 1965, but was not measured subsequently.
\item Ba, La, Ce, Pr, Nd and Sm are $\sim1$ dex overabundant; only Ba
  had been measured previously when it was also $\sim1$ dex overabundant.
\end{itemize}

The Y and Zr abundances are a mystery. Without dredge-up there
  should be no change in these elements. With dredge-up there should
  be large changes in many other elements. Bl\"ocker \&
  Sch\"onberner (\cite{Blo97}) assumed that some elements could have
  been locked in grains, but that only moves the problem to one of how the
  grains would be dissociated as the star cools.

%______________________________________________________________________________
% Figure "cohen_ew"   Comparison of Coehn abundancesa as fn of nH
%
\begin{figure*}
\epsfig{file=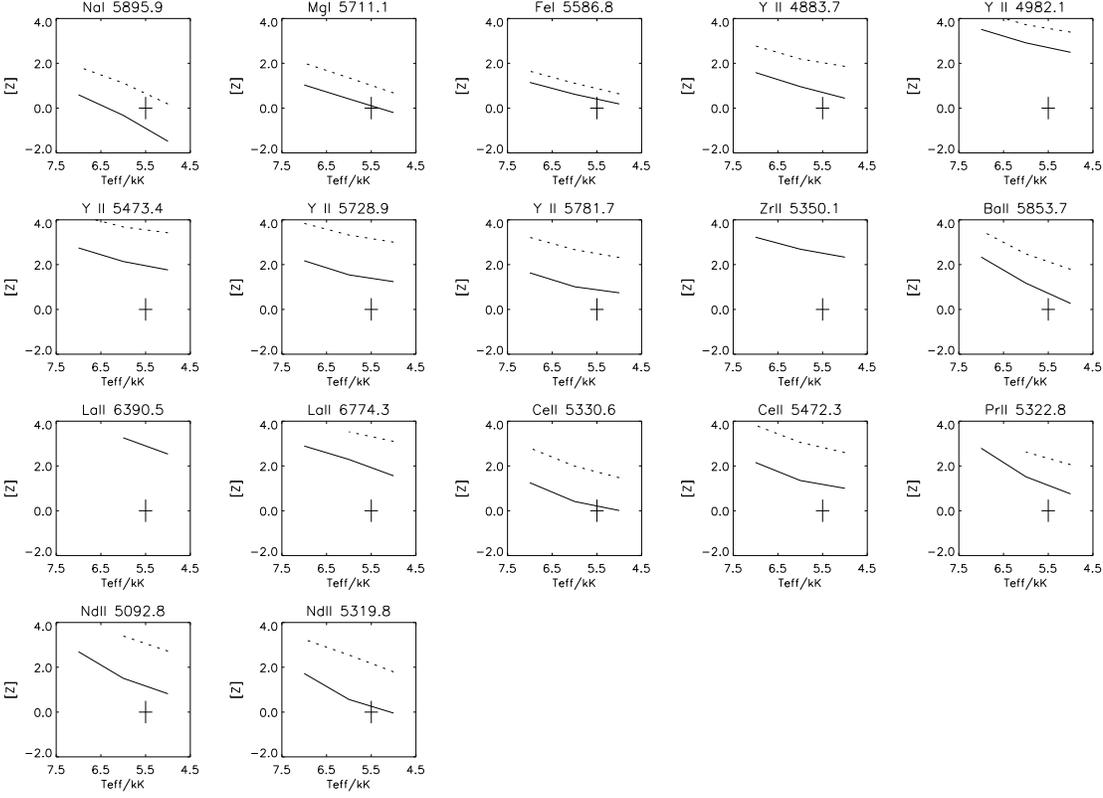,width=150mm} % ,bbllx=160,bblly=15,bburx=350,bbury=350}
 \caption[Cohen equivalent widths]{Elemental abundance excess as a
 function of \Teff\  derived from the
 equivalent widths of Cohen \& Phillips (\cite{Coh80}). Solid and dashed lines 
 refer to $\vt=10$ and 5\kmsec\ respectively. Abundances are
 averaged over the spectra obtained in 1975, 1976 and 1978
 and computed for a normal hydrogen abundance. 
 The ordinate [Z] refers to log abundance (by number)
 normalized to the solar value. The $+$ symbol represents 
 solar abundance at the \Teff\ appropriate for FG\,Sge in this epoch.
 }
 \label{f:cohen}
\end{figure*}
%_____________________________________________________________________________
% Figure "cohen_f3"   Comparison of Coehn abundancesa as fn of nH
%
\begin{figure}
\epsfig{file=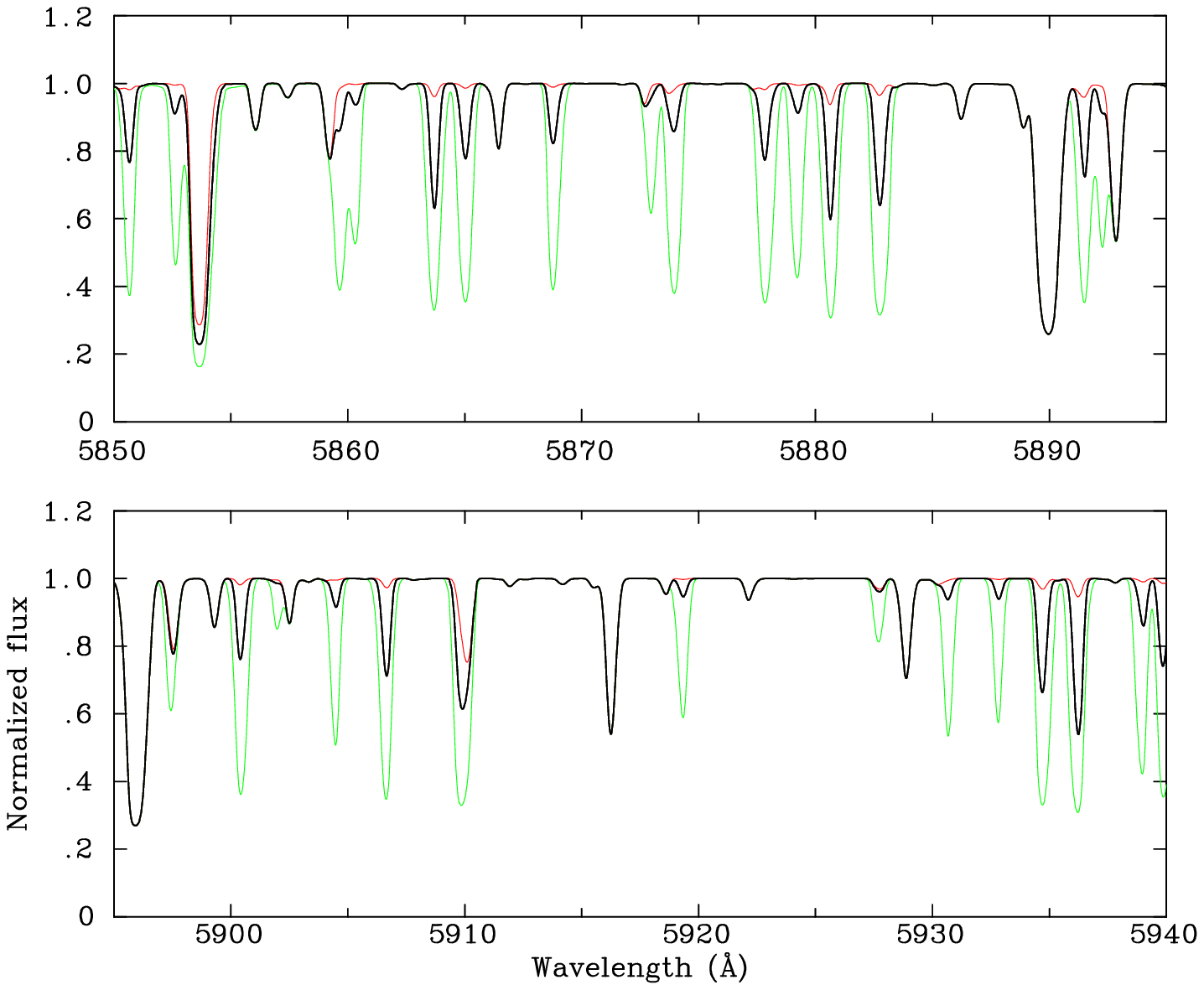,width=88mm}
 \caption[Cohen figure 3]{Synthetic spectra in the region of the NaD
 lines computed for comparison with Fig.\,3 of Cohen \& Phillips (\cite{Coh80}). 
 All spectra are calculated for a model atmosphere with $\Teff=6\,000\kelvin$, 
$\sg=0.5$, $\vt=10\kmsec$ and convolved with a 0.4\,\AA\ FWHM Gaussian. 
 Three different compositions are assumed: (1) normal
 (dotted/red), (2) s-process elements enhanced by
 1 dex (solid/black) and (3) s-process enhanced by 2 dex (dotted
 green). 
}
 \label{f:coh_f3}
\end{figure}
%_____________________________________________________________________________

\subsection{Spectral evolution in the 1970's}

Cohen \& Phillips (\cite{Coh80})  give equivalent widths for a limited number of 
strong lines measured from echelle spectra obtained on 1975 June 24, 
1976 July 7, and 1978 June 17. Using the new grid of model
atmospheres, elemental abundances were computed as a function of
\Teff, \vt\ and \nH.  Figure~\ref{f:cohen} shows 
the results for normal \nH\ only.  In line with
arguments presented elsewhere, normal \nH, $\vt=10\kmsec$ and
$\Teff=5\,000\kelvin$ are the most appropriate to use for these data. 
The main
conclusions are summarised below. There are 
uncertainties associated with the equivalent width measurements and
the atomic data used in the line formation calculations.  These are
illustrated by a spread of 2 dex in the results from five \ion{Y}{ii} lines. 
Currently the only \ion{Y}{i} line 
and two \ion{Fe}{i} lines measured by Cohen \& Phillips (\cite{Coh80}) give 
abundances which are discrepant from other lines due to the same atoms
by a further 2 dex. There is little or no evidence for a change in
composition during the interval spanned by  these data.

The individual line analyses are supported by examination of 
one echelle order of the 1975 spectrum of \object{FG\,Sge}  
(Cohen \& Phillips \cite{Coh80}, Fig. 3). Model spectra have been computed with the
abundances of s-process elements and lanthanides Sr, Y, Zr, Ba, 
La, Ce, Pr, Nd, Sm and Eu enhanced by 0, 1 and 2 dex in turn
for a range of \Teff, \nH\ and \vt. 
Figure~\ref{f:coh_f3} shows the models for normal hydrogen, 
$\Teff=5\,500\kelvin$, $\sg=0.4$ and $\vt=10\kmsec$. A comparison
with Fig.\,3 of Cohen \& Phillips (\cite{Coh80}) demonstrates that all of the principal 
features in the observed spectrum can be reproduced with an 
excess of $\sim 1$ dex in the s-process elements except, of course, 
for the interstellar component of the NaD line.  

The conclusions from the line analyses in the latter half of the
1970's include:
\begin{itemize}
\item Na, Mg and Fe were approximately normal;
\item Y and Zr were overabundant by 1--2 dex, subject to a 
      large line-to-line scatter in Y;
\item Ba, Ce, Pr and Nd were overabundant by $\sim 1$ dex;
\item La was overabundant by $\sim 2$ dex.
\end{itemize}

%______________________________________________________________________________
% Figure ``iuefits''  
%
\begin{figure}
\epsfig{file=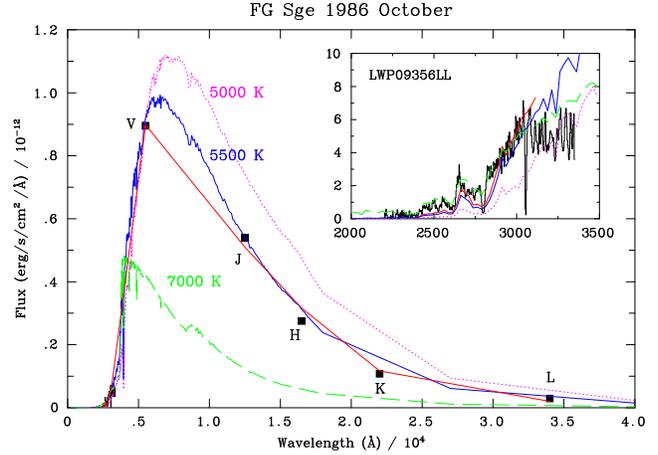,width=85mm} % ,bbllx=160,bblly=15,bburx=350,bbury=350}
 \caption[IUE fits]{Model atmosphere fits to the observed spectral
 energy distribution of FG\,Sge in 1986 October. The main panel shows
 the UV flux distribution from IUE image LWP\,09356 (histogram, lower left), 
 the visual and infrared fluxes (Montesinos et al.\ \cite{Mon90}), and a series of
 theoretical models for $\Teff= 5\,000$ (dotted, cyan), 5\,500 (solid, blue)
 and 7\,000\,K (dashed, green).
 The first two are normalised to the observations at 5500 \AA, the
 latter is the optimum fit to the UV data. The bold (red) line is the best
 fit solution from {\sc ffit} cited in Table~\ref{t:iue}. The inset
 panel shows an enlargement of the same figure in the IUE range. }
 \label{f:iue}
\end{figure}
%_____________________________________________________________________________

\begin{table*}
\caption{Effective temperatures and angular radii for FG\,Sge from
  spectral energy distributions.}
\label{t:iue}
\begin{center}
\begin{tabular}{  l c c  c c c c  c c c  }
\hline\hline\noalign{\smallskip}
Date & IUE & V &  J & H & K & L & \Teff  & $ \theta$  & $\chi_{\nu}^2$ \\
     & image & (FES)      &    &   &   &   & (kK)   & ($/10^{-10}$
rad) &  \\
\noalign{\smallskip}\hline\noalign{\smallskip}
1982 June    & LWR\,13399 & 9.4  & 7.0 & 6.8 & 6.6 & 6.6 & 5.45$\pm$0.08 &
5.15$\pm$0.05 & 0.47 \\
1986 October & LWP\,09356 & 9.1  & 7.0 & 6.6 & 6.4 & 6.1 &
5.42$\pm$0.05& 6.08$\pm$0.02 & 3.05 \\
1987 July    & LWP\,11269 & 9.2  & 7.0 & 6.7 & 6.5 & 6.2 &
5.39$\pm$0.05& 5.42$\pm$0.05 & 0.91 \\
1987 Sept    & LWP\,11585 & 9.1  & 7.0 & 6.6 & 6.5 & 6.3 &
5.30$\pm$0.04& 5.62$\pm$0.05 & 0.57 \\
1987 October & LWP\,11820 & 9.2  & 7.1 & 6.7 & 6.5 & 6.1 &
5.28$\pm$0.04& 5.36$\pm$0.05 & 0.84 \\ 
\noalign{\smallskip}\hline
\end{tabular}
\end{center}
\end{table*}

\subsection{$T_{\rm eff}$ in the 1980's}   \label{s:iue.teff}

There are three published views of the run of \Teff\
after 1980. Visual photometry provided either a  $\sim4\,500$ \kelvin\ or
$\sim5\,500$ \kelvin\ solution (van Genderen \& Gautschy \cite{vGG95}), whilst the ultraviolet
fluxes observed with IUE 
led to a still higher $\sim6\,500$ \kelvin\ solution (Montesinos et
al.\ \cite{Mon90}). The
latter are based on a very small
fraction of the total energy distribution of the star. This may
be verified by combining the UV data with the
visual and infrared data presented by the same authors (Table \ref{t:iue}), and by
comparing these with theoretical spectral energy  distributions
(SED's, Fig. \ref{f:iue}). 

Best fit model energy distributions have been obtained using a
$\chi^2$ minimization procedure ({\sc ffit}, Jeffery \cite{Jef01}).
We have used SED's from the models computed for this paper and 
have assumed $\Ebv=0.40$ (van Genderen \& Gautschy \cite{vGG95})
and a standard Galactic extinction law (Seaton \cite{Sea79}). 
Only IUE data between 2400 and 
3100\,\AA\ were used, larger data errors at $\lambda>3100$\,\AA\ 
substantially degraded the fit quality. We could not differentiate 
qualitatively between models having 1\%, 10\%, and 90\% hydrogen (by number); 
this introduces a systematic uncertainty in \Teff\
of a few tens of degrees, comparable with the formal fitting errors
quoted  in Table~\ref{t:iue}.  Random errors in \Teff\ due to other
factors (primarily data quality) are $\sim\pm100$\kelvin, and
systematic errors due to other  properties of the models may also be
 $\sim100$\kelvin.

We conclude that the high \Teff\ photometric solution of van Genderen \&
Gautschy (1995) is consistent with the ultraviolet and infrared
photometry. We suspect that the measurements of
  Montesinos et al.\ (\cite{Mon90}) were systematically in error 
  due to the inclusion of noisy IUE data  at  $\lambda>3100$\,\AA\ and 
  since they are completely inconsistent with the contemporary 
  $VHJKL$ photometry.

%______________________________________________________________________________
% Figure "halpha"   Comparison of H-normal and H-poor atmospheres
%
\begin{figure*}
\epsfig{file=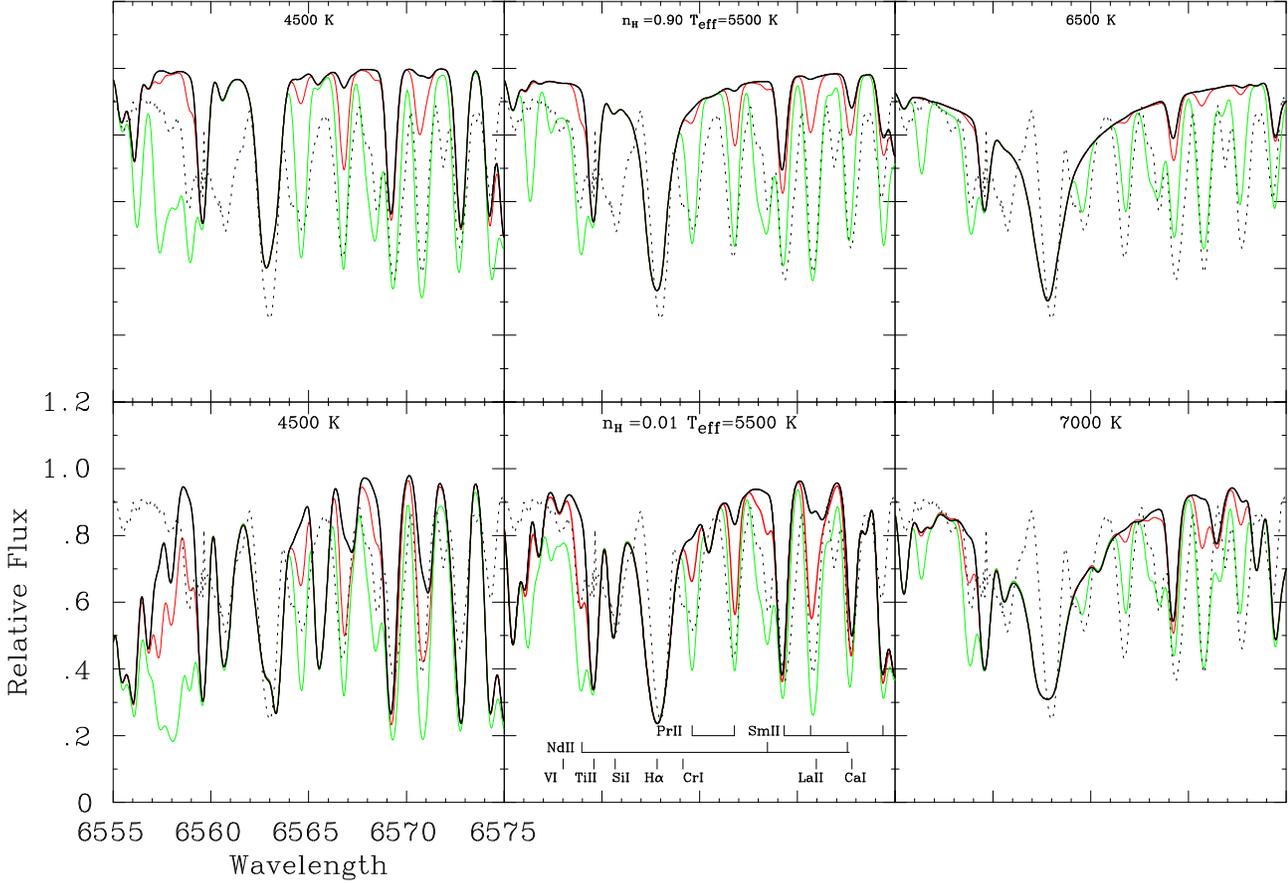,width=170mm}
 \caption[H$\alpha$]{Model spectra 
  in the region of H$\alpha$ for several values of $\Teff$ (labelled) 
  with $\log L/M=4.0$. 
  The top row shows models with normal hydrogen abundance ($n_{\rm
  H}=0.90$), and with s-process abundances increasing from normal
  (heavy black)
  through +1 dex (red) to +3 dex (green).
  The second row shows hydrogen-deficent models ($n_{\rm
  H}=0.01$), and with s-process abundances increasing from normal
  (heavy black) through +1 dex (red) to +2 dex (green).
  The dotted line is the 1994 October 21 spectrum from 
  Gonzalez et al.\ (\cite{Gon98}). The model spectra have been convolved with
  a 0.5\,\AA\ FWHM instrumental broadening function. Lines identified 
  are from Kipper \& Kipper (\cite{Kip93}).     
 }
 \label{f:halpha}
\end{figure*}
%_____________________________________________________________________________

\subsection{Did s-process abundances continue to increase into the 1990's?}

The most recent analysis of the \object{FG\,Sge} photosphere before
    it was obscured by circumstellar dust finds
    a continued increase in s-process and rare-earth elements
    (Gonzalez et al.\ \cite{Gon98}).
    Section \ref{s:gonz.etal} discusses how this conclusion was based on
    a measurement of $\Teff=6\,500\kelvin$ which both contradicts 
    the total flux distribution (Section \ref{s:iue.teff}) and implies a
    rapid contraction between 1992 and 1994. 

Prof. G. Gonzalez has kindly provided a copy of his 1994
    September 21 spectrum for comparison with our
    models. While, we have not had the resources to carry out a complete
    fine analysis of this spectrum, we have been able to test (and
    revise) various hypotheses. 
    As before, we have
    assumed $\log L/M = 4.0$ and have computed theoretical spectra 
    for \Teff\ in the range  4\,500 to 7\,000\kelvin. We have examined
    models with  $\vt=5\kmsec$ and 10 \kmsec, and three hydrogen
    abundances, $\nH=0.90, 0.10$ and 0.01. We have also considered
    models with elemental abundances for Sc, Ga, Sr, Y, Zr, Mo, Ba,
    La, Ce, Pr, Nd, Sm, Eu, Gd, Tb, Dy, Er, Yb, Lu, Hf and Pb 
    increased by 0, 1, 2 and 3 dex.

    Figure \ref{f:halpha} is plotted for direct comparison with Fig.~14 of 
    Gonzalez et al.\ (\cite{Gon98}). It shows a subset of the above
    models together with the spectrum of 1994 Sept 21 around
    H$\alpha$. The wavelength region shown includes lines of several s-process
    elements (La, Nd, Pr and Sm) as well as other species (Si, V, Cr,
    Ti).

 With normal hydrogen ($\nH=0.90$), $\Teff=6\,500\kelvin$, 
  $\log L/M=4.0$, and a 3 dex
  enhancement of s-process elements, the model does a reasonable job
  of reproducing the observed spectrum, except that the s-process lines
  are somewhat weak and the region close to H$\alpha$, especially 
  Si{\sc i}, does not fit well. A better fit to the s-process lines 
  is recovered with  $\Teff=5\,500\kelvin, \log  L/M=4.0,$ and [s/Fe] = +3 redward of
  H$\alpha$. However the ``continuum'' and the region to the
  blue are not reproduced. 

  Models with $\Teff=5\,500\kelvin$ match the measured IUE fluxes. 

 With reduced hydrogen ($\nH=0.01$), $\Teff=5\,500\kelvin$ and
  $\log L/M=4.0$, several features are already quite well reproduced with
  normal (solar) abundances.  
  Indeed, assuming a near-normal abundance for silicon, it appears 
  in this case that the observed strength of Si{\sc i} is a good
  indicator of hydrogen-deficiency. 
  Exceptions to the fit quality are at the red and blue extremes. The
  remaining features can be reproduced with $1 < {\rm [s/Fe]} < 2$.
  These lower [s/Fe] values result entirely from the reduced hydrogen
  abundance, which lowers the background opacity and immediately
  strengthens all of the absorption lines (for weak lines, 
  line strength varies as the ratio of line opacity to local
  continuous opacity). 

    Unfortunately, we are not in a position to 
    model the C$_2$ bands and hence to establish whether the carbon
    abundance is significantly different from normal. 
     The mere fact, however, that  C$_2$ is apparent indicates with certainty
     that ${\rm C/O} > 1$ and that the carbon abundance must be
    somewhat enhanced
      (see Section~\ref{s:kip.kip}).

 In summary, by adopting a more appropriate \Teff\ and gravity
 the \object{FG\,Sge} spectrum in 1992 shows:
\begin{itemize}
\item reduced hydrogen abundance,
\item a C/O ratio $ > 1$,
\item modest ($\sim 1-2$ dex) overabundances of La, Nd, Pr and Sm,
\item by extension, detections of 3 dex overabundances in other
  elements seem premature. 
\end{itemize}

A complete reanalysis of this spectrum using appropriate model atmospheres would
be rewarding.

\subsection{Summary}

Taken as a whole, this re-evaluation provides the following view 
of the evolution of \object{FG\,Sge} between 1960 and 1995:
\begin{itemize} 
\item The high \Teff\ scale (van Genderen \& Gautschy \cite{vGG95}) based on  visual photometry
  and pulsation period proxies is consistent with IUE and
  infrared photometry, and excludes a contraction/heating
  phase from the 1980s's onwards. 
\item \object{FG\,Sge} appears to have been hydrogen-rich at the start of 
  this interval, but hydrogen-poor by the end. 
\item There is little evidence for an unusual carbon abundance before
  the appearance of C$_2$ bands in the 1990's, which demand C/O $> 1$.
  However, the available data are sparse (Herbig \& Boyarchuk
  \cite{Her68}) and the C abundance in 1960 is poorly defined. 
\item In general, the abundances of s-process elements and rare-earths
  are consistently about 1 dex above the solar value. With one
  exception, the evidence for wholescale changes of up to 3 dex
  must be viewed with scepticism. A modest
  increase (up to $\sim 1.5$ dex) between 1975 and 1994 may be acceptable, 
  representing a confidence level on our estimates. 
\item The only credible
  exception is a 1 dex enhancement in Y and Zr between 1965 and
  1972. This is a mystery without dredge-up.
\end{itemize}

%___________________________________________________________________    Figure 
%  
\begin{figure*}
\sidecaption
\includegraphics[width=12cm]{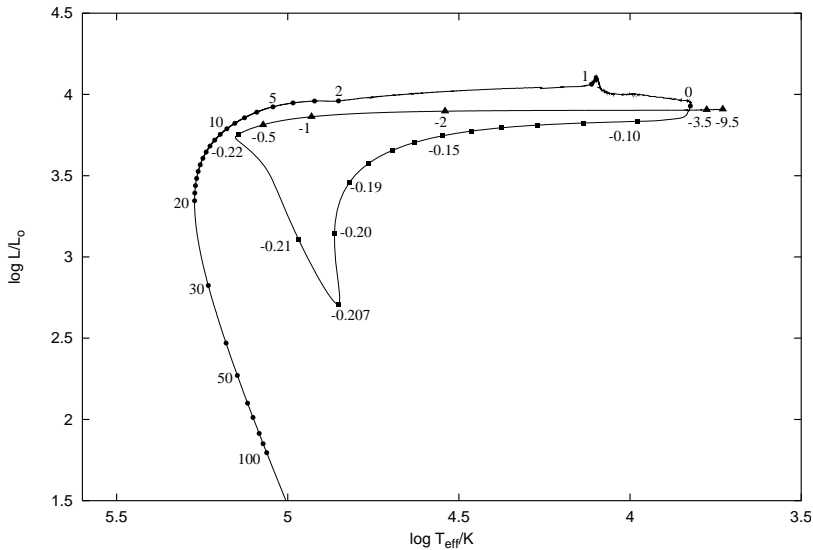}
\caption{Evolutionary track of a 0.625\Msolar\ post-AGB model suffering from a
         late thermal pulse (LTP). Symbols along the track give the time
         elapsed in units of $10^3$y, with zero set to  the minimum 
         effective temperature after the helium shell flash. Dredge-up
         commences shortly thereafter. Age $-3\,500$\,yrs 
         corresponds to the beginning of the post-AGB evolution.
         The figure is from Bl\"ocker et al.\ (\cite{Blo01b}).
        }       
\label{f:sch1}
\end{figure*}
%______________________________________________________________________________
\begin{table}
\caption{Blueward evolution of a 0.625\Msolar\ LTP model,
including principal abundances % used in Fig.~\ref{f:future} 
written as $ N_{\rm el}$ (adapted from Bl\"ocker \cite{Blo01}).
	 }
\label{t:future}
\begin{center}
\begin{tabular}{ ccccccc }
\hline\hline\noalign{\smallskip}
\Teff & \sg & Age (yr) & H & He & C & O \\
\noalign{\smallskip}\hline\noalign{\smallskip}
 6700 & 0.55 & 0    & 0.918 & 0.079 & 0.003 & 0.000 \\
 7000 & 0.63 & 128  & 0.877 & 0.119 & 0.003 & 0.001 \\
 7900 & 0.80 & 161  & 0.793 & 0.182 & 0.021 & 0.005 \\
 9000 & 1.00 & 195  & 0.657 & 0.279 & 0.052 & 0.013 \\
10000 & 1.20 & 279  & 0.577 & 0.335 & 0.071 & 0.017 \\
11200 & 1.37 & 374  & 0.461 & 0.416 & 0.099 & 0.024 \\
11900 & 1.47 & 418  & 0.321 & 0.513 & 0.134 & 0.032 \\
12300 & 1.49 & 507  & 0.181 & 0.607 & 0.170 & 0.041 \\
12600 & 1.49 & 788  & 0.162 & 0.620 & 0.176 & 0.042 \\
13800 & 1.70 & 1073 & 0.162 & 0.620 & 0.176 & 0.042 \\
15800 & 1.94 & 1098 & 0.162 & 0.620 & 0.176 & 0.042 \\
\noalign{\smallskip}\hline
\end{tabular}
\end{center}
\end{table}
%______________________________________________________________________________

\section{Evolution}                 \label{s:evol}

   The evolution of \object{FG\,Sge} is defined
   photometrically (van Genderen \& Gautschy \cite{vGG95}) since about 1890 and 
   spectroscopically since 1960. An additional upper limit of 
   $\Teff \simeq 50\,000\kelvin$
   before expansion is provided by 
   the associated planetary nebula (He 1--5: Hawley \& Miller \cite{Haw78}), 
  taking into account the weakness of the observed He\,{\sc ii} emission
      and the recombination time scale. 
   The nebula is reported by the same authors to have normal 
   abundances, although a previous
   report of anomalously high N/O coupled with $\Teff \la 
   55\,000\dots75\,000\kelvin$ should be acknowledged (Harrington \&
   Marionni \cite{Har76}). 
   These constraints may be discussed in terms of theoretical models for 
   the evolution of \object{FG\,Sge}. 

\subsection{Late thermal pulse models}

\label{s:ltp}

   The upper limit on \Teff\ defined where the helium-shell flash occurred
   during  \object{FG Sge's} 
   contraction towards a white-dwarf configuration and 
   indicates that this event happened along the horizontal part of its post-ABG evolution.
   This type of flash is normally called a `late' thermal pulse (LTP)
   as opposed to a `very late' thermal pulse (VLTP) 
   which occurs on the descending
   part of the post-AGB track (cf.\ Bl\"ocker \cite{Blo01}).
   From a theoretical point of view 
   the characteristic
   difference between both types of thermal pulse is that only for the
   VLTP does the pulse-driven convective shell penetrate into 
   hydrogen-rich layers, thereby causing complete mixing and burning of the
   hydrogen-rich envelope matter (Sch\"onberner \cite{Sch79}; Iben et al.\ \cite{Ibe83};
   Iben \cite{Ibe84}; Herwig et al.\ \cite{Her99}).  
   According to the latter authors, the final surface abundance pattern 
   following a VLTP  
   should reflect virtually that of the intershell region, viz.\ a mixture of 
   helium, carbon and oxygen, without hydrogen (see also Herwig \cite{Her01a}).

   The speed of the redward looping depends on the
   characteristic thermal/gravitational time scale of all the matter above 
   the helium shell and is a strong function of the remnant mass. Based
   on `flashing' post-AGB models of different masses Sch\"onberner \&
   Bl\"ocker (1997) estimated for \object{FG\,Sge} a mass very close
   to 0.6~M$_{\odot}$. 

   It is important to note here that in the VLTP mixing 
   and burning of the hydrogen-rich
   envelope occurs immediately during the flash and is already completed 
   {\it before} the star turns back to the red-giant region.
   On the contrary, a central star experiencing an LTP is not
   expected to show any signs of internal mixing/burning processes at its
   surface while it is expanding towards a `born-again' red giant 
   configuration. This appears to be confirmed by the normal surface
   composition found for \object{FG~Sge} after it attracted the attention
   of observers, {\it i.e.} at a time when the flash is already over. 

   In any case, the observed enrichment of the s-process elements
   need not be the result
   of the late thermal pulse (cf.\ Iben \cite{Ibe84}). The s-process 
   elements are likely to be
   the result of the past pulses when \object{FG~Sge} was still an AGB
   star. Indeed, several post-AGB stars are already known to be
   enriched in C and N, with C/O $>1$, weak iron-group elements and
   s-process elements by $\sim 1-2$ dex (Reddy et
   al.\ \cite{Red99}, \cite{Red02}, Hrivnak \& Reddy
   \cite{Hri03}). With very similar surface compositions, such stars
   are likely counterparts to
   \object{FG\,Sge} should they suffer a LTP sometime in the future.  
   
    Though the standard evolutionary models of an LTP 
    develop strong envelope convection 
    while approaching low surface temperatures, the bottom of the convection
    zone does not penetrate into nuclearly processed layers, {\it
    i.e.}\ the 3rd dredge-up does not occur until after the star 
    reaches its minimum temperature.

    This is illustrated in Fig.~\ref{f:sch1} and Table~\ref{t:future} 
    by means of the 
    0.625~M$_{\odot}$ born-again model of Bl\"ocker (\cite{Blo01b}).
    Dredge-up starts after the reddest point is reached, leading to a slow
    dilution of the hydrogen-rich envelope by helium/carbon-rich
    intershell matter, with a mean rate of $\simeq 10^{-5}$~M$_{\odot}$/yr. 
    Note that there occurs no burning of hydrogen. During the dredge-up the
    star contracts slowly towards $\Teff=10\,000\kelvin$, at a rate
    of 10~K/yr. The dredge-up is completed within about 800 years and leads
    to a hydrogen-deficient stellar surface: $X=0.05,\ Y=0.45,\
    X_{\rm C}=0.38,\ X_{\rm O}=0.12$. The model has obviously changed into
    a [WC]-type central star. A very similar result has been obtained by
    Herwig (2001) for a 0.6~M$_{\odot}$ post-AGB model.
    
    Therefore the dilemma posed by FG\,Sge is that, according to the
    associated planetary nebula, it must have experienced a
    LTP. However, LTP models predict that surface hydrogen should not
    be depleted until a deep convection zone is established, and then
    only $\sim10^4$ y after the pulse occurs. In the case of FG\,Sge, 
    this hydrogen-depletion has occurred {\it before} the star reaches
    its minimum temperature, but substantially {\it after} the time at which
    it is predicted to occur following a VLTP. 

\subsection{Double-loop models}

    An alternative and interesting model has been
    proposed by Lawlor \& MacDonald (\cite{Law03}). 
    In their calculations of the
    evolution of a post-AGB star following a 'very late thermal
    pulse', {\it ``models with low convective mixing efficiency first evolve
    quickly to the AGB, return to the blue, and then evolve more
    slowly back to the AGB for a second time before finally returning
    to the white dwarf cooling track''}. The attraction of this model
    is that it appears to provide a very satisfactory explanation for the rapid
    redward evolution of \object{V4334\,Sgr} (Duerbeck et al.\ \cite{Due02}), the
    historical appearance and subsequent contraction of \object{V605\,Aql}
    (Clayton \& de Marco \cite{Cla97})
    and the comparatively leisurely expansion of \object{FG\,Sge} as
    different phases of the same evolutionary track. 

    According to this model, \object{FG\,Sge} would be in its second return to
    the red, but
    the weaknesses of the model are several. It predicts that the
    surface of \object{FG\,Sge} should be highly-processed
    (i.e.\ hydrogen-deficient and carbon-rich) throughout its recent
    history which, we have shown, it was not. It also
    states that the star would have been a white dwarf, with
    $\Teff>100\,000\kelvin$, in the comparatively recent past;
    too hot to comply with the nebula diagnostics (cf.\ Sect.\ \ref{s:evol}),
    and possibly too old for the nebula to have survived (which it has).

    Very recently, Miller Bertolami et al. (\cite{Mil06}) have made
    an independent calculation of evolution following a very late thermal
    pulse, and again find a double-loop structure with hydrogen and
    then helium providing two phases of expansion. The interepretation
    of \object{FG\,Sge} within this framework suffers the same
    difficulties as for the  Lawlor \& MacDonald (\cite{Law03})
    models.

\section{Conclusions}

We have made an extensive review of the chemical evolution of the
surface of \object{FG\,Sge} between approximately 1960 and 1990, an
epoch during which there have been published reports of dramatic
changes in the abundances of many elements including particularly 
those created by s-process nucleosynthesis in the stellar interior. 
In several cases, these various reports have been shown to be seriously
inconsistent with one another and with stellar evolution
theory. Although it is more usual to recast the theory when
observation and theory come into conflict, the observational
inconsistencies are sufficiently strong that they must be examined closely.

Although we have not had direct access to the majority of the raw spectrograms,
we have been able to assess the reliability of the
majority of key studies, particularly those by Herbig \& Boyarchuk
(\cite{Her68}),
Langer et al.\ (\cite{Lan74}), Cohen \& Phillips (\cite{Coh80}),
 and Gonzalez et al.\ (\cite{Gon98}).
Although the first three of these report some changes in the
abundances of carbon and s-process elements, we find that the data
presented are almost entirely consistent with a constant surface composition
typical of other post-AGB stars, i.e.\ with \mbox{C/O $>1$} and s-process
abundances enhanced by $\sim 1 \mbox{-} 2$ dex. The exception that the
earliest data for Y and Zr indicate an increase from normal abundance 
to $\sim +1$ dex remains a mystery. The very dramatic increases in
s-process abundances reported in the most recent study (Gonzalez et
al.\ \cite{Gon98}) are demonstrably wrong. They are
due to the adoption of a value for \Teff\ which is 1\,000\kelvin\ too high.
A reassessment of the calibration of the photometric
evolution from ultraviolet data by Montesinos et al.\ (\cite{Mon90})
demonstrates that the ``high'' \Teff\ calibration by van Genderen \&
Gautschy (\cite{vGG95}) is most consistent with all of the photometric and
spectroscopic data. Since \object{FG\,Sge} has now maintained its
current  \Teff\ of around 5\,500\kelvin for over a decade, this may 
represent the cool limit in its redward evolution.

Supporting Iben (\cite{Ibe84}), we have argued that the current
surface of \object{FG\,Sge} does not show evidence of freshly
dredged-up material in the form of carbon or s-processed
elements. Contrary to Sch\"onberner \&
Jeffery (\cite{Sch03}), we find that 
surface hydrogen has been depleted in the recent
past ($\sim40$ y). In the 1960's it had an apparently normal H
content (cf. Fig. 6), whereas in 1994 there is evidence
of H-depletion via a reduced background opacity (cf. Fig. 12);
a conclusion based on a partial analysis of the 
high-resolution spectrum of Gonzalez et al. (\cite{Gon98}) 
not earlier at our disposal.
We do not see evidence of the very prompt 
mixing suggested by recent VLTP theory.

Stellar evolution theory predicts that, although a
convective envelope does form as the star cools following a late
thermal pulse, a deep convective
envelope capable of dredging up processed material will not form until
some time after the star has completed its redward
evolution. Therefore, the observed hydrogen depletion, suggesting
prompt envelope mixing following a VLTP, contradicts the age derived
from the  associated planetary nebula, which suggests an LTP. The
question is then how well the physics of the LTP and VLTP are
understood, and whether the demarcation between them is as concrete as
current models suggest.

\object{FG\,Sge} is one of those very rare objects which demonstrates
stellar evolution ``while-you-wait''. As a very evolved star, its
surface already shows some material which has been created by
nucleosynthesis in the deep interior. 
Astero-archaeology is the science of sifting through such
material for clues to the past evolution of the star. In this case, 
most of the material was brought to the surface 
when the star had a deep convective envelope on the asymptotic
giant branch. However, it seems that \object{FG\,Sge} has recently 
entered a new period of excavation, which has recently 
brought fresh helium to the surface. Other elements will 
be added to this within the next hundred or so years, when it will 
be possible to observe fresh material being brought to the surface 
as the star begins to dig up its own past. Continuing observations 
will provide key tests for stellar evolution models, particularly for
models of late and very late thermal pulses.

%
% ____________________________________________________________ Acknowledgments
%
\section*{Acknowledgements}

This research was supported by a Northern Ireland 
Department of Culture, Arts and Leisure
grant to the Armagh Observatory and by visitor grants from the 
Astrophysikalisches Institut Potsdam.
This research has made extensive use of the SIMBAD database,
operated at CDS, Strasbourg, France, of NASA's Astrophysics Data System,
and of the INES System, developed by the ESA IUE Project at VILSPA.

The authors are grateful to: Ms E. Stephenson, a summer student sponsored by the 
Nuffield Foundation, for assistance with remeasurement of the IUE
data, to Prof G. Gonzalez for providing a copy of the 1994 spectrum of
FG\,Sge, to Dr T. Bl\"ocker for details of his evolutionary models,
to Dr R. Stancliffe for teaching one of us about the ramifications of
the Very Late Thermal Pulse. They are indebted to the referee, 
Dr M. Asplund, for his careful critiques of this paper
which resulted in a reappraisal of their original conclusions.

\appendix

\section{Herbig \& Boyarchuk abundance measurements} \label{a:hb68}

\begin{table*}
\caption[HB abundances]{Elemental abundances in \object{FG\,Sge}:
a) relative to \object{$\alpha$\,Cyg} (Herbig \& Boyarchuk \cite{Her68})
(cols 2 --  6),
b) mean values (1960 -- 1965, excluding 1960 Sept.) 
normalized such that the 
combined abundances of Si, Ti, Cr and Fe are the same in 
\object{FG\,Sge} and \object{$\alpha$ Cyg} 
(Albayrak \cite{Alb00}) (col 8), 
c) the same mean values relative to
\object{$\alpha$ Cyg} 
(col 7) and the \object{Sun} (col 9), 
and d)  elemental abundances for  \object{$\alpha$\,Cyg} (Albayrak \cite{Alb00})
and the \object{Sun} (cols 10 -- 11). Abundances [X] are given as
logarithm of relative abundance by number normalized to H = 12.00 
}
\label{t:hbabs}
\small
\begin{center}
\begin{tabular}{  c ccccc  ccc  cc  }
\hline\hline\noalign{\smallskip}
Elem & \multicolumn{5}{c}{FG\,Sge: [X]$-\alpha$\,Cyg} &
\multicolumn{3}{c}{FG\,Sge: Mean} & $\alpha$\,Cyg  &  Sun \\
     & July  & Sept  & Sept  & Aug   & May   & [X]$-\alpha$\,Cyg &  [X]  & [X]$-$Sun
& & \\
     & 1960  & 1960  & 1962  & 1963  & 1965  &   & &  &  &    \\
\noalign{\smallskip}\hline\noalign{\smallskip}
H    & +0.50 & +0.53 & +0.23 & +0.55 & +0.43 & +0.43 &12.43 &+0.43 &12.00 &12.00 \\
He   & +1.48 &$-0.90$& --    & --    & +0.06 & +0.77 &11.62 &+0.69 &10.85 &[10.93] \\
C    & +0.74 &$-0.18$& +0.91 & +0.83 & +1.09 & +0.89 & 9.83 &+1.44 & 8.94 & 8.39 \\
O    & --    & --    & +0.25 & +0.74 & +0.45 & +0.48 & 9.11 &+0.45 & 8.63 & 8.66 \\
Mg   & +0.29 & +0.85 & +0.10 & +0.48 & +0.06 & +0.23 & 7.74 &+0.21 & 7.51 & 7.53 \\
Si   & +0.04 &$-0.17$& +0.28 & +0.45 & +0.19 & +0.24 & 7.64 &+0.13 & 7.40 & 7.51 \\
Ca   & +0.60 & +1.61 &$-0.25$&$-0.19$&$-0.37$&--0.05 & 6.74 &+0.43 & 6.79 & 6.31 \\
Sc   & --    & --    & +0.36 & +0.41 & +0.16 & +0.31 & 3.26 &+0.21 & 2.95 & 3.05 \\
Ti   & +0.05 & +0.15 &$-0.07$&$-0.13$&$-0.03$&--0.05 & 4.96 &+0.05 & 5.00 & 4.90 \\
V    & --    & --    & +0.11 & +0.49 & +0.26 & +0.29 & 4.41 &+0.41 & 4.12 & 4.00 \\
Cr   & +0.17 & --    &$-0.01$&$-0.14$& +0.03 &  0.01 & 5.73 &+0.09 & 5.72 & 5.64 \\
Mn   & +0.29 & --    & +0.15 &$-0.12$& --    & +0.11 & 5.71 &+0.32 & 5.60 & 5.39 \\
Fe   &$-0.28$& +0.03 &$-0.21$&$-0.17$&$-0.13$&--0.20 & 7.37 &--0.08& 7.57 & 7.45 \\
Co   & +0.92 & --    & +0.44 & +0.61 & --    & +0.66 & --   & --   &  --  & 4.92 \\
Ni   & --    & --    & +0.13 & +0.23 & +0.22 & +0.19 & 6.43 &+0.20 & 6.24 & 6.23 \\
Sr   & +1.06 & --    & +0.57 & +0.82 & +0.56 & +0.75 & 4.03 &+1.11 & 3.28 & 2.92 \\
Zr   & --    & --    & +0.05 & +0.32 & +0.33 & +0.23 & 3.64 &+1.05 & 3.41 & 2.59 \\
Ba   & --    & --    & +0.22 & +0.32 & +0.38 & +0.31 & 3.19 &+1.02 & 2.88 & 2.17 \\
Eu   & --    & --    &$-0.5$ & +0.1  &$-0.2$ &--0.20 & 2.28 &+1.76 & 2.48 & 0.52 \\
\noalign{\smallskip}\hline
\end{tabular}
\end{center}
\end{table*}

\noindent In order to make the abundance measurements for FG\,Sge presented by Herbig \&
Boyarchuk (\cite{Her68}) more accessible, they are presented
in Table~\ref{t:hbabs} in the more conventional form of the logarithm
of the abundance by number relative to $\alpha$\,Cyg. 
Adopting a recent measurement of abundances in $\alpha$\,Cyg (Albayrak
\cite{Alb00}), the
mean Herbig \& Boyarchuk (\cite{Her68}) abundances in FG\,Sge are also shown relative to the Sun. 
Note that Albayrak (\cite{Alb00}) gives the C{\sc ii} abundance in
$\alpha$\,Cyg as 8.94 from one line only. The C{\sc i} abundance is
8.21. 
Since the  Herbig \& Boyarchuk
measurements were based on C{\sc ii} lines, we have used the C{\sc
 ii} value for $\alpha$\,Cyg in Table~\ref{t:hbabs}. 
The mean overabundance of C relative to the Sun between 1960 and 1965 in 
\object{FG\,Sge} would be +0.4 dex rather than +1.1 dex if the 
C{\sc i} value were adopted for $\alpha$\,Cyg.

% _________________________________________________________________ references
%

\end{document}